\documentclass[a4paper,11pt]{article}

\pdfoutput=1

\usepackage{jinstpub}
\usepackage{amsmath}
\usepackage{mathtools} % multlined
\usepackage{siunitx}[=v2] % Typesetting values with units 
\usepackage{upgreek} % To get upright greek symbols
\usepackage[shortcuts]{extdash} % Unbreakable dashes
\usepackage{booktabs} % Better tables
\usepackage{multirow} % Better tables
\usepackage{lineno}

\modulolinenumbers[5]
\graphicspath{{figures/}}
\newcommand{\fig}[1]{figure~\ref{fig:#1}}
\newcommand{\Fig}[1]{Figure~\ref{fig:#1}}
\newcommand{\eq}[1]{equation~(\ref{eq:#1})}

\newcommand{\tab}[1]{table~\ref{tab:#1}}

\newcommand{\sect}[1]{section~\ref{sec:#1}}
\newcommand{\Sect}[1]{Section~\ref{sec:#1}}

\DeclareSIUnit\electron{\ensuremath{\textnormal{e}}}
\DeclareSIUnit\hits{\ensuremath{\textnormal{hits}}}
\DeclareSIUnit\bit{\ensuremath{\textnormal{bit}}}

\DeclareMathOperator\erf{erf}

\makeatletter \newcommand*\fsize{\dimexpr\f@size pt\relax} \makeatother

\title{Timing performance of the Timepix4 front-end}

\author[a,1]{K.~Heijhoff,\note{Corresponding author.}}
\author[a]{K.~Akiba,}
\author[b]{R.~Ballabriga,}
\author[a]{M.~van~Beuzekom,}
\author[b]{M.~Campbell,}
\author[a]{A.P.~Colijn,}
\author[a]{M.~Fransen,}
\author[a]{R.~Geertsema,}
\author[a]{V.~Gromov,}
\author[b]{and X.~Llopart Cudie}

\affiliation[a]{Nikhef, Science Park 105, 1098 XG Amsterdam, the Netherlands}
\affiliation[b]{CERN, 1211 Geneva 23, Switzerland}

\emailAdd{k.heijhoff@nikhef.nl}

\abstract{A characterisation of the Timepix4 pixel front-end with a strong focus on timing performance is presented. Externally generated test pulses were used to probe the per-pixel time-to-digital converter (TDC) and measure the time-bin sizes by precisely controlling the test-pulse arrival time in steps of \SI{10}{\pico\second}. The results indicate that the TDC can achieve a time resolution of \SI{60}{\pico\second}, provided that a calibration is performed to compensate for frequency variation in the voltage controlled oscillators of the pixel TDCs. The internal clock distribution system of Timepix4 was used to control the arrival time of internally generated analog test pulses in steps of about \SI{20}{\pico\second}. The analog test pulse mechanism injects a controlled amount of charge directly into the analog front-end (AFE) of the pixel, and was used to measure the time resolution as a function of signal charge, independently of the TDC. It was shown that for the default configuration, the AFE time resolution in the hole-collecting mode is limited to \SI{105}{\pico\second}. However, this can be improved up to about \SI{60}{\pico\second} by increasing the preamplifier bias-current at the cost of increased power dissipation. For the electron-collecting mode, an AFE time resolution of \SI{47}{\pico\second} was measured for a bare Timepix4 device at a signal charge of \SI{21}{\kilo\electron}. It was observed that additional input capacitance from a bonded sensor reduces this figure to \SI{62}{\pico\second}.}


\begin{document}
	\maketitle
	\flushbottom
	% \linenumbers
	
	\section{Introduction}
	The High Luminosity Large Hadron Collider (HL-LHC) \cite{Apollinari:2017} is an upgrade to the existing Large Hadron Collider that will increase the potential for new discoveries by increasing the integrated luminosity by a factor of ten. As a consequence, the number of concurrent collisions per bunch crossing (referred to as pile-up) will increase, making it more difficult for particle physics experiments to distinguish between collisions based on the spatial measurements of collision products by the detectors closest to the interaction point. To prevent large amounts of pile-up, the instantaneous luminosity will be kept at a constant level for the majority of the time during which collisions take place (a few hours typically); normally the instantaneous luminosity peaks at the start and decays over time, but luminosity levelling prevents the initial peak while keeping the average high enough to reach the desired integrated luminosity.	Despite luminosity levelling, the tracking of decay products based on spatial measurements will likely be insufficient to assign them to the correct primary vertex because the reconstructed tracks will not have the required spatial resolution to distinguish between spatially overlapping vertices. A potential solution that is currently being pursued in the particle physics community is the incorporation of time measurements in the detectors to aid in the reconstruction of tracks and primary vertices\===a method typically referred to as 4D~tracking~\cite{Cartiglia:2017, Sadrozinski:2018}. The time resolution that can be achieved with pixel detectors is therefore of great interest.
	
	The time resolution of a pixel detector partly depends on the physical processes that happen in the sensor material in which charged particles deposit part of their energy, which generates a detectable electronic signal \cite{Riegler:2017}. Another important contribution to the time resolution is from the front-end electronics that amplifies and discriminates the sensor signals. Lastly, the time at which the signal is discriminated is converted to a digital representation by a time-to-digital converter~(TDC) which also contributes to the time resolution. New sensor technologies are being developed in order to achieve the time resolution required for 4D tracking at the HL-LHC, and the results are promising \cite{Lai:2020, Paternoster:2021}. However, there is currently no full-scale pixel readout ASIC with a front-end that is fast enough to benefit from the intrinsic time resolution provided by these new fast-sensor technologies. In a previous study \cite{Heijhoff:2021} it has been shown that 3D \cite{Parker:1997} and thin planar detectors bonded to the Timepix3 pixel ASIC \cite{Poikela:2014} have a time resolution that is limited by the pixel front-end. 
	
	Timepix4~\cite{Llopart:2022} is the latest ASIC in the Medipix family \cite{Ballabriga:2020}. It is the successor to Timepix3 and Timepix~\cite{Llopart:2007}, which have been used across a wide range of applications, partly owing to their ability to do per-pixel time measurements.\footnote{Chronologically, Timepix2 \cite{Wong:2020} was developed after Timepix3 as a successor to Timepix in order to meet a demand pertaining to applications that do not necessitate the added complexity of data-driven readout associated with Timepix3. The Timepix chips are named according to the collaboration by which they are developed: Timepix3 was developed by the Medipix3 collaboration that developed the Medipix3 ASIC. At the time, the Medipix2 collaboration did not develop a second version of Timepix alongside Medipix2.} In this study the timing performance of the new Timepix4 ASIC is characterised by means of test signals. The pixel TDC is studied by externally generated signals that are routed to the digital front-ends of pixels near the bottom and top peripheries, and the time resolution of the analog front-end is characterised by internally generated test pulses. The TDC measurements are performed using only bare ASICs (without a bonded sensor). The analog front-end measurements are performed also with devices that are bonded to \SI{300}{\micro\meter} planar silicon p-on-n sensors which provide a realistic input capacitance to the preamplifier. Two sequential revisions of the Timepix4 ASIC, versions 1 and 2, are tested.
	
	This paper is structured as follows. \Sect{timepix4} gives a general introduction of Timepix4 and explains its relevant features. \Sect{digitalFrontEnd} covers the timing performance of the digital front-end, and presents calibration measurements that are required in \sect{analogFrontEnd}, which present the analog front-end measurements and their results. \Sect{conclusion} contains the conclusion.

	\section{The Timepix4 pixel ASIC} \label{sec:timepix4}
	Timepix4 has been produced in a \SI{65}{\nano\meter} CMOS technology. It has a matrix that consists of $448\times512$~pixels, which is a factor \num{3.5} bigger than Timepix3. Furthermore, the maximum hit rate has been improved by a factor 8 to a value of \SI{358}{{\mega\hits}\per{\square{\centi\meter}}\per{\second}}. Most importantly, concerning the subject of this paper, it offers an improved analog timing performance, and also has a more precise TDC featuring time bins of~\SI{195}{\pico\second} compared to the \SI{1.56}{\nano\second} bin size in Timepix3. Like its predecessor, Timepix4 can measure the time of arrival (ToA) and time over threshold (ToT) of each hit simultaneously. The latter is a surrogate measure for the amount of charge in a signal, which can be used to measure particle energy. It can also be used to correct systematic errors in the time measurement that depend on signal size, as will be discussed in \sect{timewalk}.

	\subsection{Pixel front-end} \label{sec:timepix4PixelFrontEnd}
\begin{figure}[htbp]
		\centering
		\includegraphics{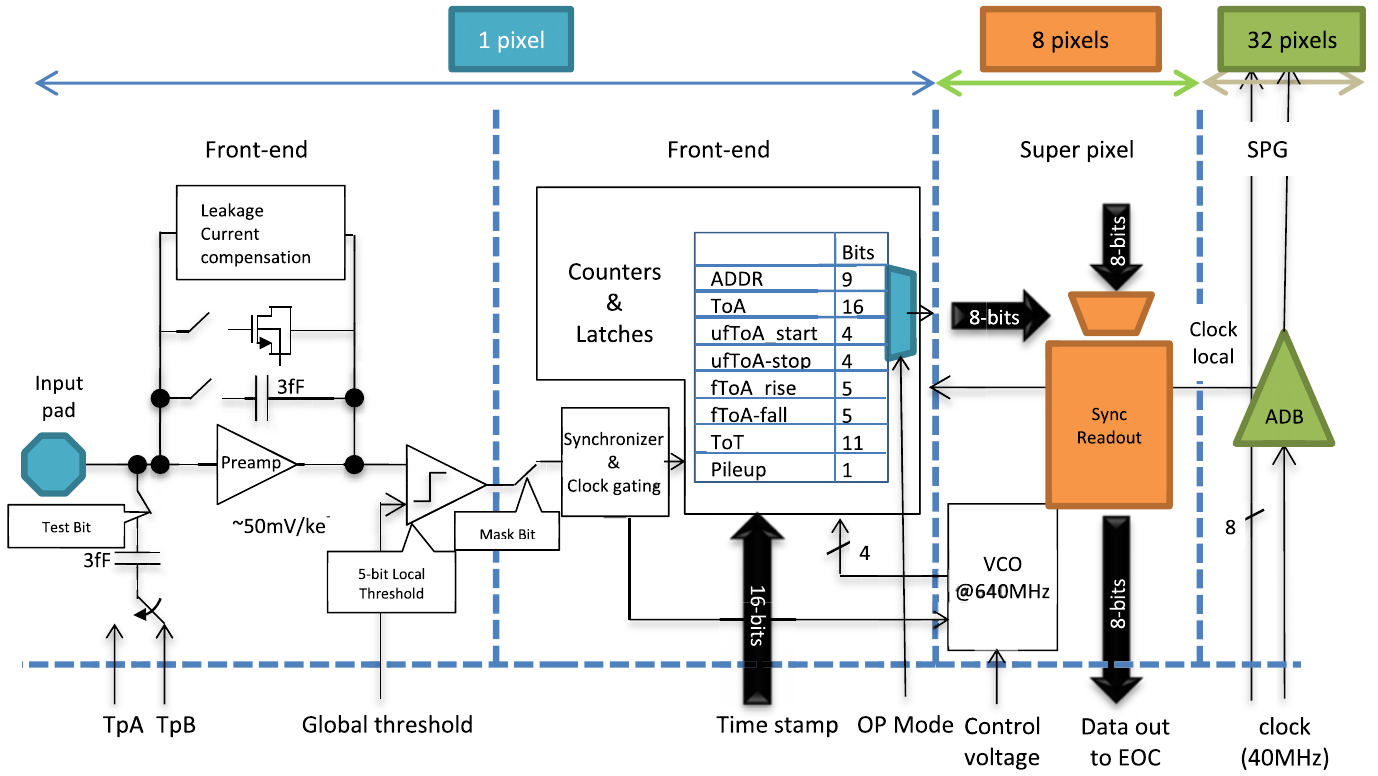}%
		\caption{Schematic diagram of the Timepix4 front-end \cite{Llopart:2022}. The front-end of a single pixel is divided into an analog and a digital part. A superpixel consists of two by four pixels and contains a \SI{640}{\mega\hertz} oscillator which is used for precise time measurements. A superpixel group (SPG) consists of four superpixels and contains two adjustable delay buffers (ADBs) that distribute the system clock along columns of pixels (only one is shown).}
		\label{fig:tpx4FrontEnd}
	\end{figure}

	\Fig{tpx4FrontEnd} shows a schematic diagram of the Timepix4 front-end. As in Timepix3~\cite{Gaspari:2014}, the charge sensitive preamplifier in the analog front-end of Timepix4 is based on the Krummenacher scheme~\cite{Krummenacher:1991}. It compensates for leakage current and it can process positive as well as negative input signals to work with both hole- and electron-collecting sensors. The preamplifier has a roughly linear relationship between input charge and ToT because the feedback capacitor, onto which the signal current at the input pad is integrated, is discharged at a constant rate. The dynamic range of the ToT measurement can be increased by enabling the low-gain mode. This mode lowers the preamplifier gain by adding an extra capacitor in parallel to the main feedback capacitance. For hole-collecting sensors, the dynamic range can be increased further by the adaptive-gain mode, which adds a MOS gate capacitance to the feedback circuit. In this study only the default high-gain mode is considered.
	
	% However, for positive polarity signals, from hole-collecting sensors, the time resolution is inherently limited because for these signals there is an upper bound to the current at which the first amplifier stage discharges the capacitive load at its output (i.e. at a certain amount of input charge, the analog front-end enters a slew-rate limited regime where an increase in the input charge does not increase the slope of the preamplifier signal at the output anymore).

	Although the analog front-end can process positive as well as negative input signals, the time resolution is expected to be better for electron-collecting sensors. For hole-collecting sensors, at a certain amount of input charge, the analog front-end enters a slew-rate limited regime where a further increase in the input charge does not increase the slope of the preamplifier signal at the output anymore~\cite{Ballabriga2022}. For positive polarity signals, there is an upper bound to the current with which the first amplifier stage can discharge the capacitive load at its output. As a consequence, the time resolution is inherently limited for hole-collecting sensors.
	
	The test-pulse circuit at the preamplifier input allows for the injection of a controlled amount of charge. The input is connected to the voltage references TpA and TpB through a capacitor in an alternating manner; each time the voltage switches, a current pulse is injected into the front-end. The duration of this injected current signal is negligible compared to the rise time of the preamplifier, and the signal can therefore be regarded as having the shape of a delta function.
	
	The discriminator after the preamplifier uses a global (chip-wide) threshold voltage. To compensate for pixel-to-pixel baseline variations, each pixel also has a five-bit local threshold setting, which is referred to as a trim DAC in this paper.

	\Fig{tpx4FrontEnd} also contains a table listing the counters and latches involved in the various time measurements of each hit. In this figure, the ToA is the timestamp corresponding to the \SI{40}{\mega\hertz} clock, which in this paper is more specifically referred to as the coarse ToA (cToA). Likewise, the ToT is the number of \SI{40}{\mega\hertz} clock cycles that the preamplifier output is above the threshold value. The so-called fine time of arrival of the rising and falling edges of the preamplifier signal, fToA-rise and fToA-fall, correspond to measurements performed with a \SI{640}{\mega\hertz} voltage-controlled oscillator (VCO) which is shared by a group of two by four pixels referred to as a superpixel. In this study only the fToA-rise counter is used, and it is simply referred to as the fToA. The even more precise ultra-fast time of arrival codes, ufToA-start and ufToA-stop, capture the phase of the VCO when the discriminator fires (start) and when the first subsequent \SI{40}{\mega\hertz} clock edge arrives (stop). The ufToA-start code is only relevant for hits that arrive when the VCO has already been activated by another hit in the same superpixel, which cannot happen with the measurement method used in this study. Therefore, only the ufToA-stop code is used, and for brevity it is simply referred to as the ufToA code, or simply the ufToA when talking about its decoded value. The next section explains the time measurement and these variables in more detail.
	
	In \sect{timingPerformance} the test-pulse circuit is used to measure the time resolution of the analog front-end as a function of input charge. In this study the reference-clock distribution system (\sect{refClockDist}) is used to control the clock phase with respect to the arrival time of the internally generated test pulses. As the clock is delayed, the time bin in which the test pulses arrive will change, and due to the noise in the analog front-end, this transition from one bin to another is not instantaneous: With each increment of the clock delay, the number of test pulses arriving in one time bin gradually decreases as they end up in the next adjacent time bin. The number of hits as a function of the clock delay therefore takes on the shape of an s-curve, which is the cumulative ToA distribution of the test pulses from which the time resolution of the analog front-end can be determined.
	
	Timepix4 offers the possibility to use the digital front-end of certain pixels to timestamp external signals. This can be useful when working in conjunction with other detectors (to provide reference signals for example). In \sect{tdcCharacterisation} this feature is used to study the TDC (which is of the same type for all pixels) by generating external signals with a controlled arrival time with respect to the reference clock. In \sect{adbCalibration} this feature is used to calibrate the reference-clock distribution, which is necessary for the analog test-pulse measurements in \sect{analogFrontEnd}.
	
	\subsection{Time measurement in Timepix4} \label{sec:tpx4TimeMeasurement}
\begin{figure}[htbp]
		\centering
		\includegraphics{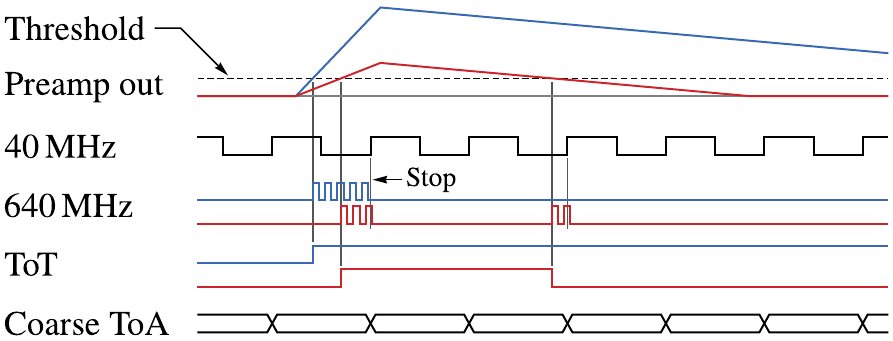}%
		\caption{Diagram of the time measurement in Timepix4 for two hits with a different signal amplitude. In Timepix4 the \SI{640}{\mega\hertz} clock is also activated on the falling edge of the discriminator in order to measure the ToT with a granularity of \SI{1.56}{\nano\second}. Figure adapted from \cite{Heijhoff:2020}.}
		\label{fig:tpx4Timing}
	\end{figure}

	The time-to-digital conversion in Timepix4 can be roughly divided into three parts, of which the first two are similar to Timepix3 (\fig{tpx4Timing}). First the coarse ToA is determined as the \SI{40}{\mega\hertz} clock cycle in which the preamplifier output goes over threshold and activates the discriminator. Secondly, the discriminator activates the VCO, and by counting its number of oscillations until the next rising edge of the \SI{40}{\mega\hertz} clock, which defines the fine ToA, the timestamp granularity can be improved to \SI{1.56}{\nano\second}. In Timepix4 the VCO is also activated when the preamplifier goes below threshold in order to achieve a more precise ToT measurement. Normally, this feature is not necessary as the uncertainty in the ToT measurement is dominated by the voltage noise on the preamplifier output due to the shallow threshold crossing of the falling edge. However, if the discharge-current setting is increased, the threshold crossing improves, and the timestamp granularity due to the \SI{40}{\mega\hertz} clock can become the dominating factor. This feature is mainly aimed at high flux applications, where it might be necessary to increase the discharge current of the feedback capacitor, and thereby shorten the preamplifier output signal to prevent signal pile-up. 
	
\begin{figure}[htbp]
		\centering
		\includegraphics{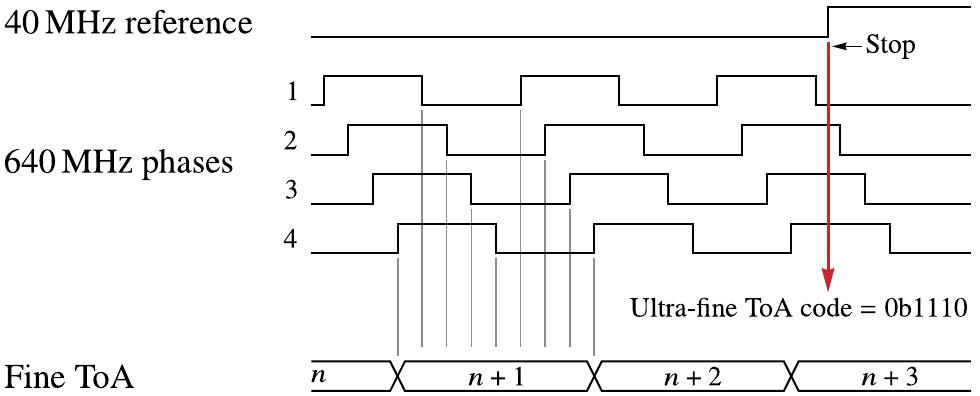}%
		\caption{Diagram showing the ultra-fine ToA measurement in Timepix4. The four phase shifted \SI{640}{\mega\hertz} clocks divide the \SI{1.56}{\nano\second} period into eight time bins of \SI{195}{\pico\second}.}
		\label{fig:uftoaDiagram}
	\end{figure}

	\Fig{uftoaDiagram} shows the third part of the time-to-digital conversion in Timepix4. The ToA measurement is further refined by having four phase shifted copies of the \SI{640}{\mega\hertz} clock. These phase shifted clocks define eight time bins within a single \SI{1.56}{\nano\second} period. The states of these four clocks are latched on the subsequent rising edge of the \SI{40}{\mega\hertz} clock after the signal has crossed threshold. This defines the four-bit ultra-fine ToA code, which is used to refine the time measurement to~\SI{195}{\pico\second}.
	
	The common control voltage of the VCOs, which determines their frequency, is normally generated in the centre periphery by a phase-locked loop (PLL) which is synchronised to the \SI{40}{\mega\hertz} reference clock. However, the second iteration of the chip (Timepix4v1) suffers from a problem in the modelling of the control-voltage dependence of the VCO frequency, which has led to a frequency that is too high.\footnote{The first iteration (Timepix4v0) suffers from the same problem.} The supply voltage of the periphery PLLs can be lowered independently to achieve the target frequency, which is necessary for the correct transmission of data. The supply voltage of the superpixel VCOs, however, cannot be set independently, and the control voltage that is generated by the periphery PLLs is too high. The control voltage is therefore configured to be taken from a dedicated DAC that is set to its lowest value to lower the frequency as much as possible. The resulting frequency is still too high, resulting in small time bins, but this does not negatively affect the front-end operation otherwise.

	\subsection{Reference-clock distribution} \label{sec:refClockDist}
\begin{figure}[htbp]
		\centering
		\includegraphics[width=120mm]{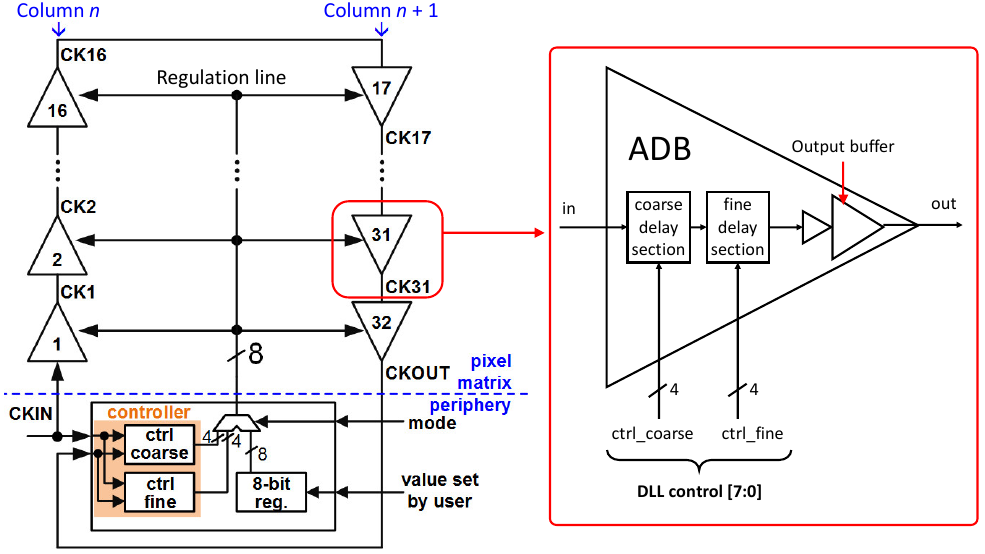}%
		\caption{Schematic of the digital delay-locked loop that distributes the \SI{40}{\mega\hertz} reference clock along a double-column. Figure adapted from \cite{Llopart:2019}.}
		\label{fig:dllSchematic}
	\end{figure}

	The \SI{40}{\mega\hertz} reference clock is distributed along each double-column structure by means of a digital delay-locked loop (DLL) in order to achieve a well-defined clock phase at the pixels with a target skew of less than \SI{100}{\pico\second}~\cite{Llopart:2019}. A schematic of the DLL structure is shown in \fig{dllSchematic}. The clock propagates away from the centre periphery along the columns to the outermost super-pixel groups and back. In each super-pixel group, the clock is buffered in both directions by an adjustable delay buffer (ADB). During normal operation, the controller, which is located in the centre periphery, regulates the delay of each of the \num{32} ADBs to \SI{781}{\pico\second} in order to achieve a total delay that is equal to one clock cycle of \SI{25}{\nano\second}. It is also possible, however, to override all controllers and set a chip-wide DLL control code to set the delay manually. Furthermore, all ADBs can be bypassed individually to prevent malfunctioning or completely defective ADBs from affecting entire double-columns. The manual control of the ADBs also allows for accurate control of the clock phase, which is demonstrated in \sect{adbCalibration}, and used in \sect{timingPerformance} to measure the analog front-end time resolution.
	
	\section{Digital front-end measurements} \label{sec:digitalFrontEnd}
	In this section the superpixel TDC is characterised using a similar method that was used to determine the time-bin sizes of Timepix3~\cite{Zappon:2015, Heijhoff:2021}. \Sect{measurementSetup} describes the measurement setup, and \sect{tdcCharacterisation} presents the results. In \sect{tdcResolution} the expected TDC resolution is determined from the measurement results. In \sect{adbCalibration} the clock distribution system is calibrated in preparation of the analog front-end characterisation described in the next section.

	\subsection{Measurement setup} \label{sec:measurementSetup}
	The digital pixel inputs of Timepix4 can be used to timestamp up to four external signals. Each half of the chip has two digital pixel inputs, and each input is routed to a different set of double-column structures within that half. These double-column structures are typically referred to by their respective end-of-column (EoC) blocks. Within each half, one input is routed to all even-numbered EoCs, and the other input goes to all odd-numbered EoCs. The external signals are routed to the digital front-ends of all pixels in the first superpixel of each EoC. These signals can be enabled on a per-superpixel basis, and pixels can be masked to prevent receiving multiple hits per superpixel. The measurements described here are performed with a single non-masked pixel in each superpixel.
	
	%The Timepix4 devices are controlled and read out using version four of the Speedy PIxel Detector Readout (SPIDR4), which is the successor of SPIDR3~\cite{Visser:2015, Heijden:2017}. \Fig{spidr4} shows the SPIDR4 control board with a Timepix4 carrier board containing a bonded device. The external signals are synchronised by means of a \SI{10}{\mega\hertz} clock from which the SPIDR4 system generates a synchronised \SI{40}{\mega\hertz} clock for the Timepix4 device. The external signal inputs are routed directly to the Timepix4 device.
	
\begin{figure}[htbp]
		\centering
		\includegraphics{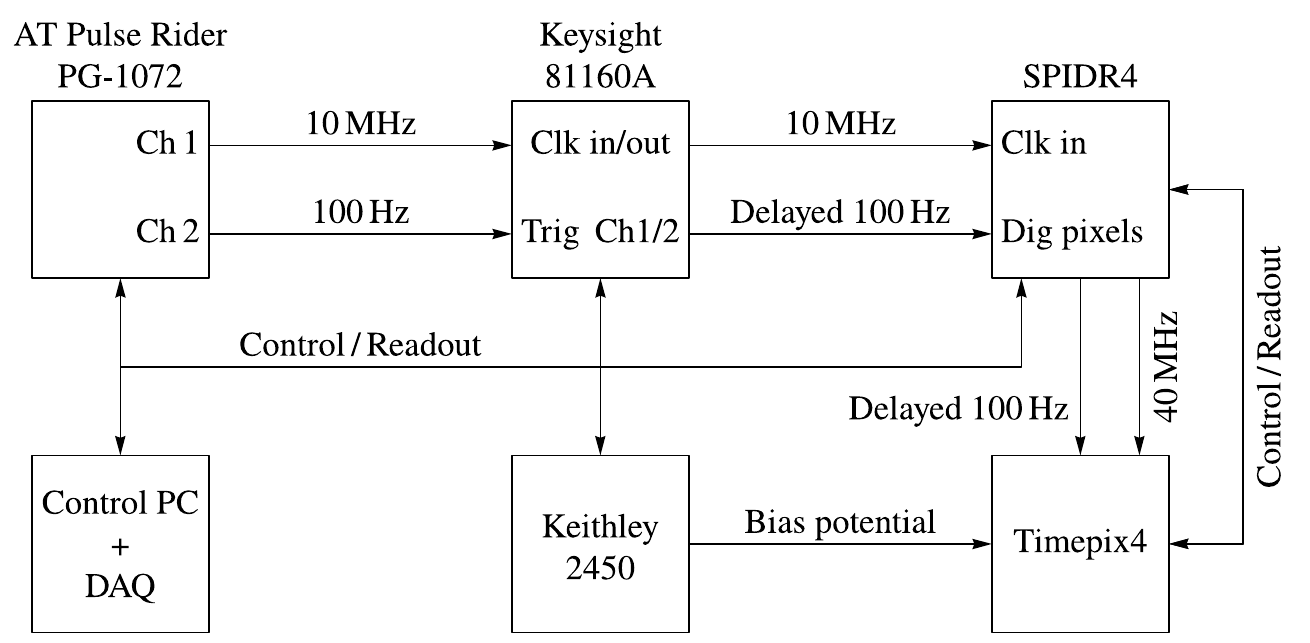}%
		\caption{Diagram of the measurement setup.}
		\label{fig:measurementSetup}
	\end{figure}

\begin{figure}[htbp]
		\centering
		\includegraphics{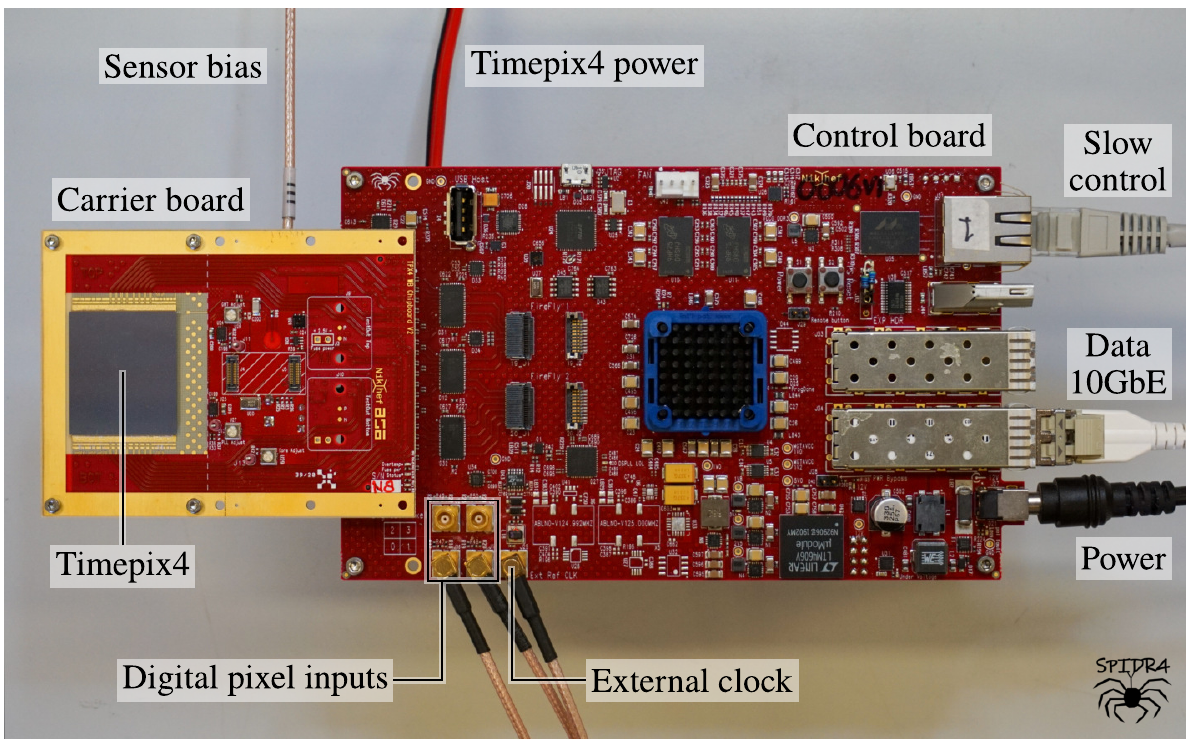}%
		\caption{A SPIDR4 control board with a Timepix4 carrier board containing a Timepix4v1 with a \SI{300}{\micro\meter} sensor. The Timepix4 power cable is directly connected to the carrier board.}
		\label{fig:spidr4}
	\end{figure}

	\Fig{measurementSetup} shows a diagram of the measurement setup that is used for the digital pixel measurements. The external test pulses are generated by two pulse generators:\footnote{The AT Pulse Rider was added after it was observed that the internal trigger of the Keysight is not sufficiently synchronised with its \SI{10}{\mega\hertz} clock. The internal trigger suffers from a drift of about \SI{0.1}{\nano\second} per hour of elapsed real time (the exact value depends on the configuration) with respect to the \SI{10}{\mega\hertz} clock.} The first pulse generator provides a \SI{10}{\mega\hertz} clock and a synchronised \SI{100}{\hertz} trigger signal to a second pulse generator, which is used to generate square pulses with a width of \SI{1}{\micro\second} that are phase shifted by means of a configurable trigger delay. The signals are then fed into the Timepix4 readout system (SPIDR4, the successor of SPIDR3~\cite{Visser:2015, Heijden:2017}) which generates a synchronised \SI{40}{\mega\hertz} reference clock for Timepix4, and passes on the test pulses to one digital pixel input of each half of the chip. \Fig{spidr4} shows the SPIDR4 control board with a Timepix4 carrier board containing a Timepix4v1 bonded to a \SI{300}{\micro\meter} p-on-n sensor. The setup allows for precise control of the test-pulse arrival time in steps of \SI{10}{\pico\second} within the \SI{25}{\nano\second} clock period of the reference clock. 
	%which generates a synchronised \SI{40}{\mega\hertz} reference clock for Timepix4, and passes on the test pulses to one digital pixel input of each half of the chip.
	
	\subsection{Superpixel TDC characterisation}\label{sec:tdcCharacterisation}
\begin{figure}[p]
		\centering
		\lineskip=5mm%
		\includegraphics{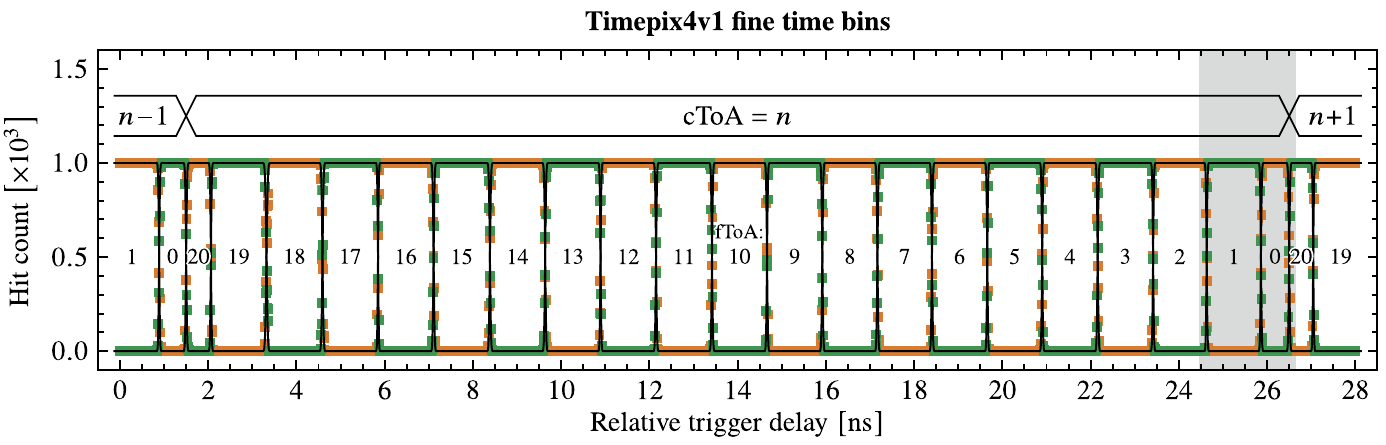}\\*[0pt]%
		\includegraphics{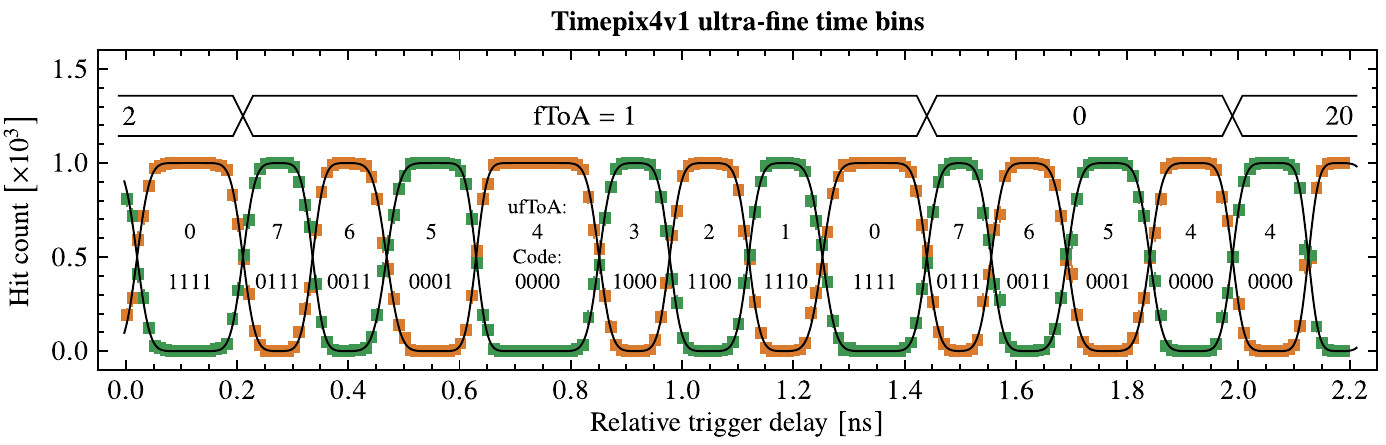}\\*[0pt]%
		\includegraphics{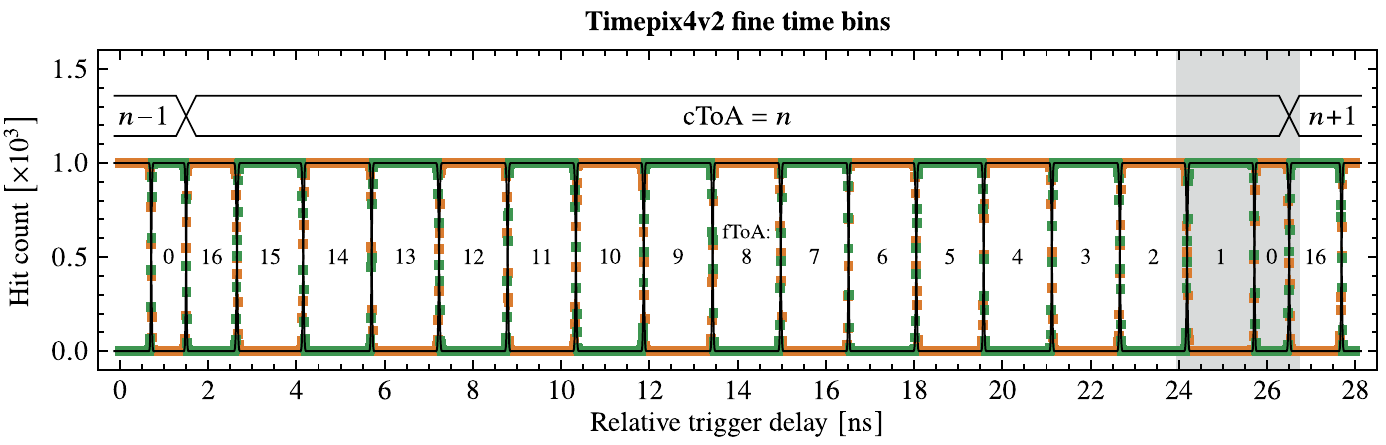}\\*[0pt]%
		\includegraphics{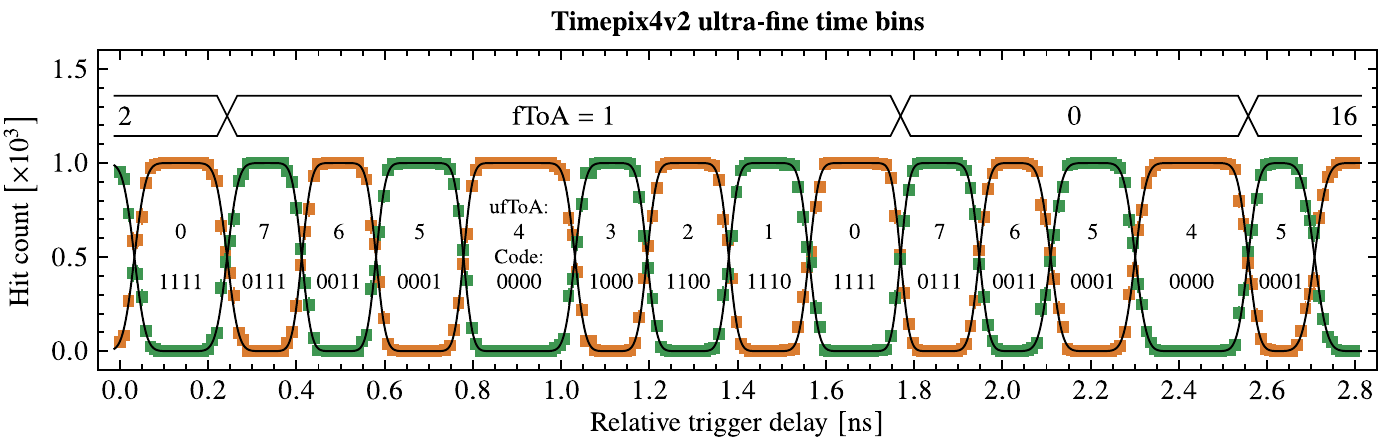}%
		\caption{Number of hits in alternating bins as seen by a single pixel of a Timepix4v1 device (top two plots) and a Timepix4v2 device (bottom two plots) as a function of trigger delay for both types of time bins. The two datasets in each plot (orange and green points) correspond to the parity of an arbitrary sequential number that has been assigned to each time bin. Timepix4v1 has smaller time bins due to a design problem of the VCO, which has been fixed in Timepix4v2.}
		\label{fig:delayScanResults}
	\end{figure}

	The TDC is characterised by scanning the test-pulse arrival time through the \SI{25}{\nano\second} reference-clock period in steps of \SI{10}{\pico\second} and recording the resulting timestamps. \Fig{delayScanResults} shows the measurement results from a single pixel for both the fine and ultra-fine time bins. It can be seen that the VCO frequency of Timepix4v1 is too high, resulting in \num{21} fine time bins. Timepix4v2 has a VCO frequency much closer to the design value, resulting in \num{17} fine bins. The first fine bins (fToA \num{0}) of both devices are about half the size of the others by design: The fToA counter is incremented by the VCO clock with the largest phase delay as was illustrated in \fig{uftoaDiagram}. The size of the last fine bin (fToA \num{20} for version 1 and fToA \num{16} for version 2) depends on the VCO frequency. The $\SI{17}{\pico\second}$ jitter on the edges is dominated by the pulse generator. The difference between the ultra-fine bin sizes is clearly visible for both devices, especially for ufToA binary codes $0000$ and $1111$. The number of hits that arrive in or after\footnote{This is a subtle detail, but it is important in ensuring that the data are well-described by an error function when the time bins are small because the hits might be distributed over more than two time bins.} a certain time bin can be modelled as
	\begin{equation}
		\label{eq:hitCountVsTriggerDelay}
		f\!\left(t_{\textnormal{delay}},\,t_n,\,\sigma_{\textnormal{j}}\right) = 
		\frac{N_{\textnormal{tp}}}{2} 
			\left[1 + \erf\!\left(\frac{t_{\textnormal{delay}} - t_n}{\sqrt{2}\,\sigma_{\textnormal{j}}}\right)\right]
		\, ,
	\end{equation}
	where $t_{\textnormal{delay}}$ is the trigger delay, $t_n$ is the left-edge location of the $n$\=/th bin where $n$ is an arbitrary sequential number assigned to the bins, $\sigma_{\textnormal{j}}$ is the jitter of the measured arrival time of the pulses with respect to the reference clock, and $N_{\textnormal{tp}}$ is the total number of pulses per measurement. In order to determine the time-bin sizes, \eq{hitCountVsTriggerDelay} is fitted to the data, and the bin sizes are then determined as
	\begin{equation}
		w_n = t_{n+1} - t_n \, .
	\end{equation}

\begin{figure}[p]
		\centering
		\includegraphics{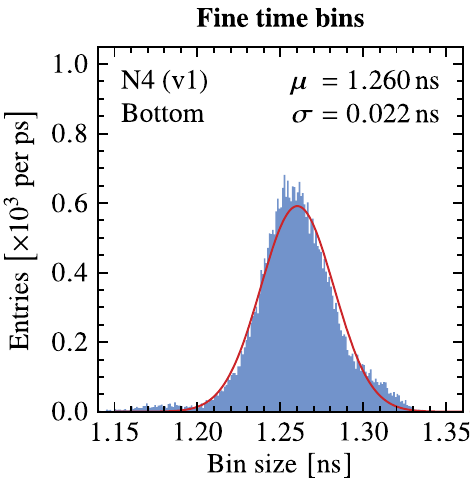}%
		\hskip3mm%
		\includegraphics{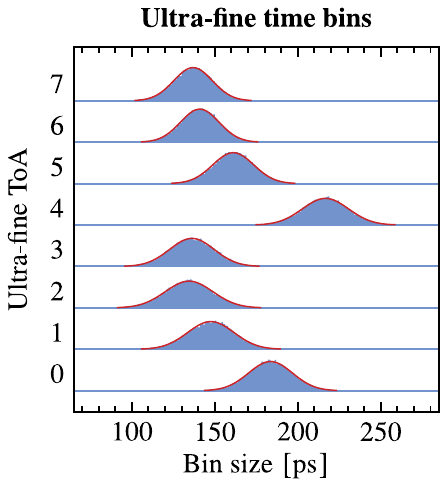}%
		\hskip3mm%
		\includegraphics{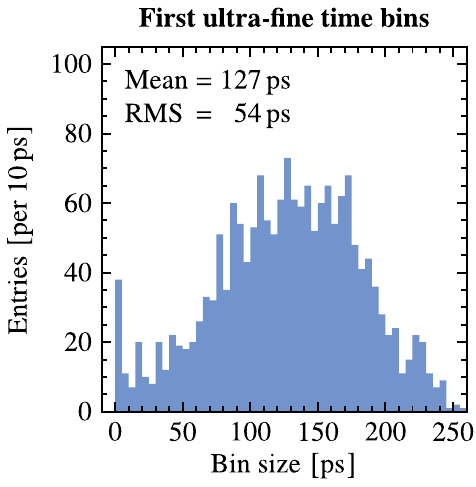}%
		\caption{Distribution of fine time bin sizes the central fine time bins with $0<\textnormal{fToA}<20$~(left), the ultra-fine time bins~(centre), and the first bins before a rising edge of the \SI{40}{\mega\hertz} reference clock~(right) for the bottom half of a Timepix4v1 device.}
		\label{fig:binSizeDistributionsV1}
	\end{figure}

	\Fig{binSizeDistributionsV1} shows the fit results for the bottom half of a Timepix4v1 device. The mean bin size of \SI{1.26}{\nano\second} is smaller than the design value of \SI{1.56}{\nano\second} due to the problem with the VCO frequency of Timepix4v1 as was explained in \sect{tpx4TimeMeasurement}. The data indicates that the VCOs are running at a mean frequency of \SI{794}{\mega\hertz} with a standard deviation of \SI{1.7}{\percent} over the superpixels. The majority of superpixels have 21 fine time bins, except for a small fraction of \SI{0.7}{\percent} that have 20 or 22 fine bins. It is also observed that the ultra-fine time bins exhibit a clear structure in size. Furthermore, time bins that are located immediately before a rising edge of the \SI{40}{\mega\hertz} reference clock show a relatively large pixel-to-pixel variation in size.
	
\begin{figure}[p]
		\centering
		\lineskip=3mm%
		\includegraphics{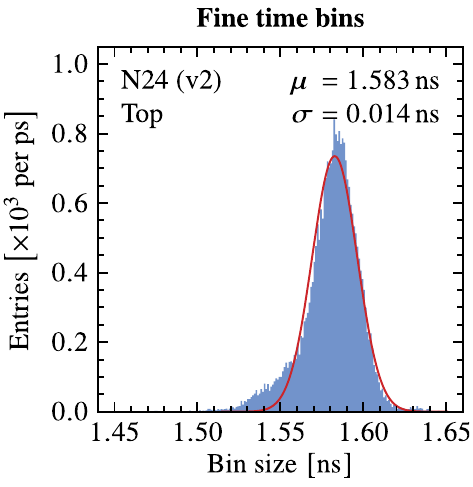}%
		\hskip3mm%
		\includegraphics{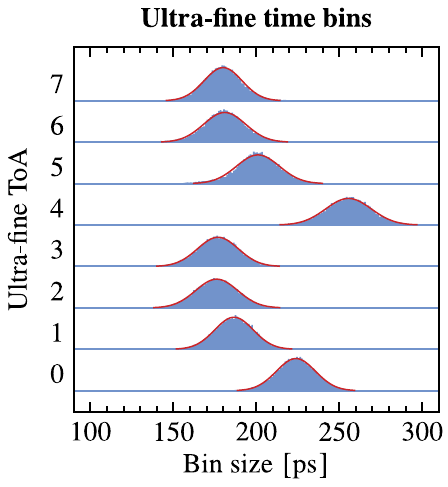}%
		\hskip3mm%
		\includegraphics{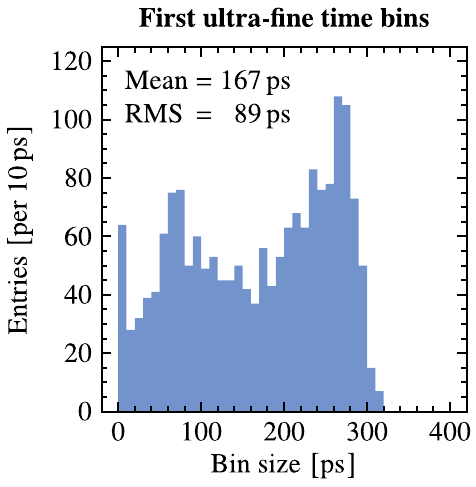}\\*[0pt]%
		\includegraphics{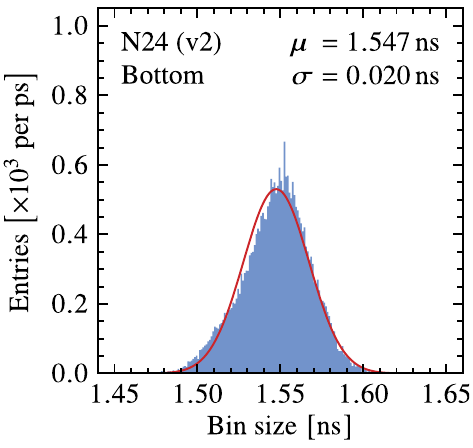}%
		\hskip3mm%
		\includegraphics{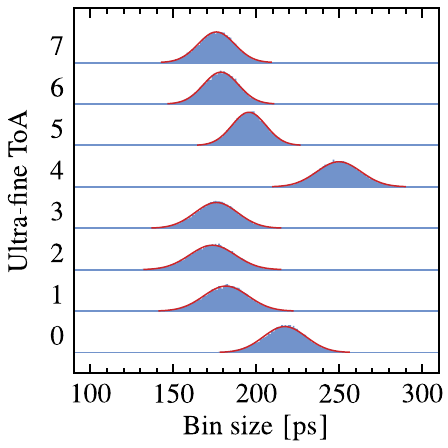}%
		\hskip3mm%
		\includegraphics{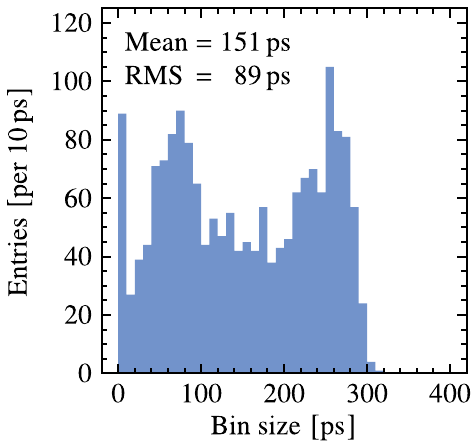}%
		\caption{Distribution of fine time bin sizes the central fine time bins with $0<\textnormal{fToA}<16$~(left), the ultra-fine time bins~(centre), and the first bins before a rising edge of the \SI{40}{\mega\hertz} reference clock~(right) for the top and bottom halves of a Timepix4v2 device.}
		\label{fig:binSizeDistributionsV2}
	\end{figure}

	\Fig{binSizeDistributionsV2} shows the results for both matrix halves of a Timepix4v2 device. A difference of \SI{2}{\percent} is observed in the mean bin size between the bottom and top halves, which could be due to a difference in the ground potential. It was checked that the variation in bin size is not correlated to the variation in the VCO control voltage due to the distribution over the EoCs. The overall bin size of \SI{1.565(17)}{\nano\second} corresponds to a VCO frequency of \SI{639(7)}{\mega\hertz}, which is in agreement with the design value. The ultra-fine bins show the same structure as observed in Timepix4v1. It is also observed that the ultra-fine time bins that are located directly in front of a rising edge of the \SI{40}{\mega\hertz} clock vary in size from almost 0 up to about \SI{300}{\pico\second}. This variation will add to the relative time offsets between pixels, and degrade the overall time resolution if not corrected. The contribution to the total time resolution can be approximated by the bin size RMS of \SI{89}{\pico\second}.
	
	The increased size of ufToA bins 0 and 4 can be understood by considering the VCO, which consists of a chain of four \SI{195}{\pico\second} delay cells whose outputs are the four clock phases that were shown in \fig{uftoaDiagram}. The first delay cell can be seen as an AND gate with a \SI{195}{\pico\second} delay; one input is the OR function of the eight discriminators in the superpixel to activate the VCO, and the other input is the inverted output of the last delay cell. The delay of this inversion was not completely accounted for in the design, and has led to an extra delay between the edges of the fourth VCO clock phase and their corresponding edges in the first clock phase, resulting in the increased bin size for ufToA 0 and 4. The difference between them is likely due to a difference in the rise and fall times of the inversion. The remaining structure is not observed in post-layout simulation, but the similarity between the two devices suggests that it is not due to process variation.
	
	\subsection{TDC time resolution}\label{sec:tdcResolution}
	In a real application of Timepix4, the hit-time of each pixel can be reconstructed as
	\begin{equation}
		\label{eq:timeReconstruction}
		t_{\textnormal{hit}}=\left(\textnormal{cToA} - \frac{\alpha}{16}\,\textnormal{fToA} - \frac{\alpha}{128}\,\Delta\textnormal{ufToA} + \alpha \beta\right) \SI{25}{\nano\second} + \Delta t_{\textnormal{pixel}}
		\, ,
	\end{equation}
	where $\Delta\textnormal{ufToA}=\textnormal{ufToA}-4$ for hits arriving when the superpixel VCO has not yet been activated by earlier hits,\footnote{As detailed in \sect{timepix4PixelFrontEnd}, this study is restricted to primary hits for which the ufToA-start code is invariably equal to $0000$, which translates to a bin number of 4. The relationship between the ufToA code and its corresponding time bin number was shown in \fig{delayScanResults}.} $\alpha$ is a correction factor that compensates for variation in the VCO frequency, $\beta$ compensates the non-uniformity of the ultra-fine bins, and $\Delta t_{\textnormal{pixel}}$ is a per-pixel term that corrects time offsets between pixels originating from differences in clock phase due to the reference-clock distribution. The latter also compensates for more subtle timing differences that originate from mechanisms such as variation in the capacitive loading of traces connecting the pixel front-ends to the superpixel VCO, as was observed in Timepix3~\cite{Heijhoff:2020, Heijhoff:2021}.
	
	To study the TDC resolution, the measured bin sizes are used to construct the expected time-residual distribution (with respect to an imaginary perfect time reference) under the assumptions that \eq{timeReconstruction} is applied and that all time offsets between pixels have been calibrated out. Initially, three types of VCO frequency corrections are considered:
	\begin{equation}
		\label{eq:frequencyCorrections}
		\begin{alignedat}{3}
			& \textnormal{(i)}  && \textnormal{\enspace chip-wide:}       	&& \quad \alpha \rightarrow \alpha_{\textnormal{chip}} ,\\*[-3pt]
			& \textnormal{(ii)} && \textnormal{\enspace per matrix half:} 	&& \quad \alpha \rightarrow \alpha_{\textnormal{half}} ,\\*[-3pt]
			& \textnormal{(iii)}  && \textnormal{\enspace per VCO:}          && \quad \alpha \rightarrow \alpha_{\textnormal{\textsc vco}} ,
		\end{alignedat}
	\end{equation}
	where the bin structure is ignored ($\beta\rightarrow0$) for all cases because its overall contribution is negligible at the current stage. For a complete pixel matrix, the number of correction factors are \num{1}, \num{2}, and \num{29e3} (the number of superpixels), respectively. An additional correction is considered in order to demonstrate the impact of the bin structure on the TDC resolution:
	\begin{equation}
		\label{eq:binStructureCorrections}
		\begin{alignedat}{3}
			& \textnormal{(iv)}   && \textnormal{\enspace per VCO and ufToA:}   	&& \quad \alpha \rightarrow \alpha_{\textnormal{\textsc vco}} \enspace \textnormal{\&} \enspace \beta \rightarrow \beta_{\textnormal{ufToA}} ,\\*[-3pt]
			% & \textnormal{(vi)}  && \textnormal{\enspace per bin:}       	&& \quad \beta \rightarrow \beta_{\textnormal{pixel, fToA, ufToA}}.
		\end{alignedat}
	\end{equation}
	which introduces \num{8} additional parameters with \num{7} degrees of freedom.\footnote{It has been assumed that all pixel offsets are calibrated out which amounts to the constraint that~${\sum_n \beta_n=0}$.}
	
\begin{figure}[htbp]
		\centering
		\includegraphics{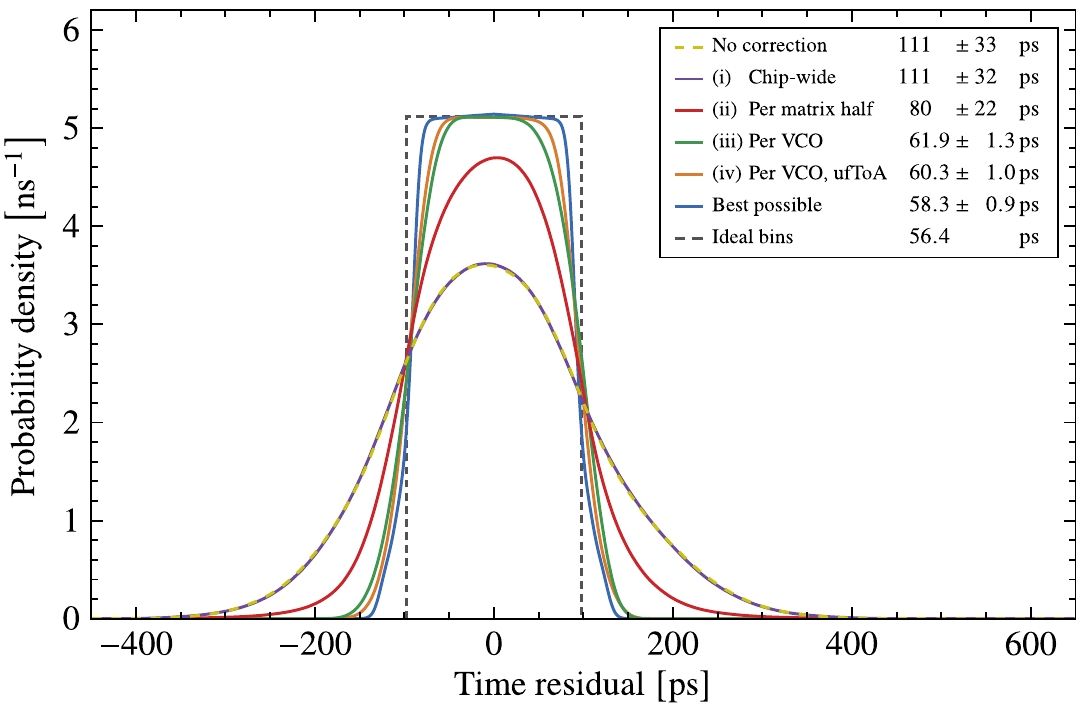}%
		\caption{Expected time-residual distributions of the TDC based on the measured bin size structure of the Timepix4v2 device for various correction methods to compensate for VCO frequency variation and the ultra-fine bin structure. For each case the overall TDC resolution is presented in the legend, where the error indicates the RMS value over the pixels.}
		\label{fig:tdcTimeResiduals}
	\end{figure}

	The correction parameters are determined by minimising the standard deviation of the resulting time-residual distribution. The results are shown in \fig{tdcTimeResiduals}. The best possible TDC resolution that can be achieved with the nominal TDC is \SI{56.4}{\pico\second}.\footnote{This figure is defined by the variance of a rectangular distribution, $\sigma^2=w^2/12$, with $w=\SI{25}{\nano\second}/128$.} However, the observed bin structure limits the best possible resolution to \SI{58.3(9)}{\pico\second}, where the uncertainty is taken as the RMS value over the pixels. When no correction is applied ($\alpha\rightarrow0$ and $\beta\rightarrow0$), a TDC resolution of \SI{111(33)}{\pico\second} is observed. Introducing a chip-wide correction factor (i) only has a very minor impact on the TDC resolution (less than \SI{1}{\percent}) because the mean VCO frequency is already very close to the design value of \SI{640}{\mega\hertz}. Correcting the VCO frequency of both matrix halves individually (ii) gives a more pronounced improvement, resulting in a resolution of \SI{80(22)}{\pico\second}. Further improvement can be achieved by taking into account the frequency of each individual VCO (iii), resulting in a resolution of \SI{61.9(13)}{\pico\second}. Incorporating the bin structure (iv) gives an additional improvement of \SI{2.6}{\percent}. Lastly, it is noted that the observed size variation in the first and last ultra-fine bins has not been taken into account in i--iv, and that doing so results in an improvement of \num{0.1} to \SI{1}{\percent}.
	
\begin{figure}[htbp]
		\centering
		\includegraphics{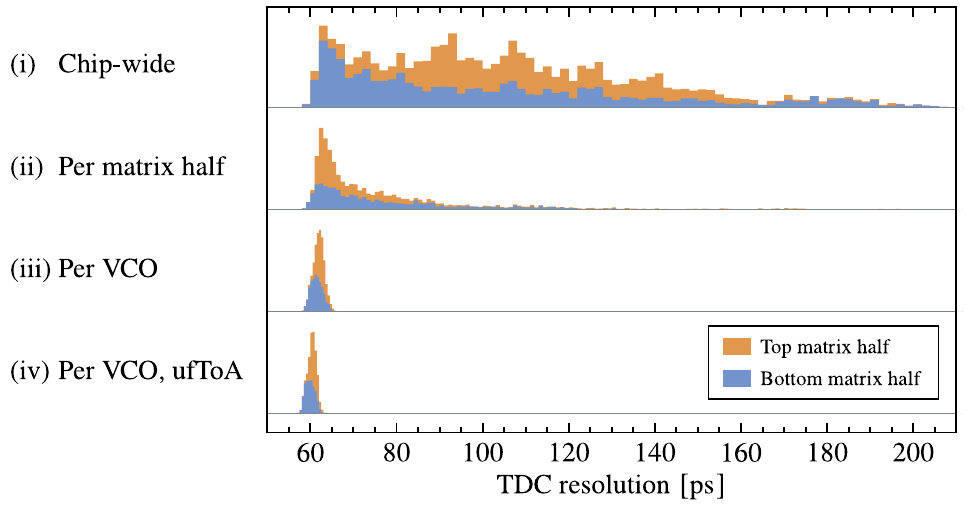}%
		\caption{Distributions of the expected TDC time resolution over the pixels for three different methods of VCO frequency corrections.}
		\label{fig:tdcResolution}
	\end{figure}

	\Fig{tdcResolution} shows how the TDC resolution is distributed over the pixels. It can be seen that methods i--ii both result in distributions with long tails towards worse TDC resolutions. For methods iii and iv it can be seen that the top half of the pixel matrix has a slightly worse resolution than the bottom half, which is due to the observed difference in the mean VCO frequency as was shown in \fig{binSizeDistributionsV2}.
	
	\subsection{Adjustable delay buffer calibration}\label{sec:adbCalibration}
	In this study the reference-clock distribution (\sect{refClockDist}) is used to control the clock phase with respect to the internally generated analog test pulses.
	The precision to which the clock phase can be controlled depends on the number of ADBs that are enabled in the chain. For the measurements described in this paper, only the first four (out of 32) ADBs are enabled. For this configuration all pixels see the same change in clock phase when the DLL control code is changed because they receive their clock from the last 16 buffers in the chain. The DLL control code determines how many coarse- and fine-delay elements are enabled within each ADB. The lowest four bits of the control code are used to enable up to 15 fine elements, and the highest four bits are used to enable up to 14 coarse elements. The four enabled ADBs are characterised by performing the same measurement as described in the previous section for different control codes. The clock phase is observed as a shift in the bin edges $t_n$ of \eq{hitCountVsTriggerDelay}.
	
\begin{figure}[htbp]
		\centering
		\includegraphics{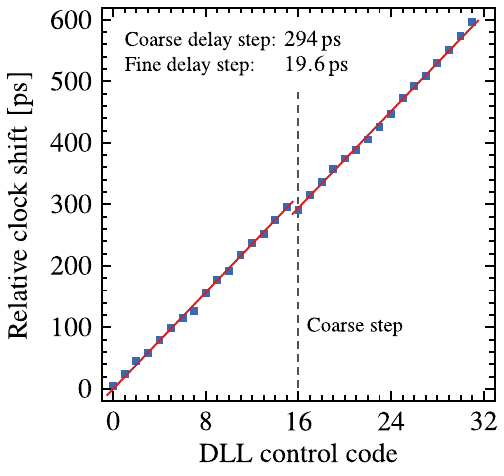}%
		\hskip4mm%
		\includegraphics{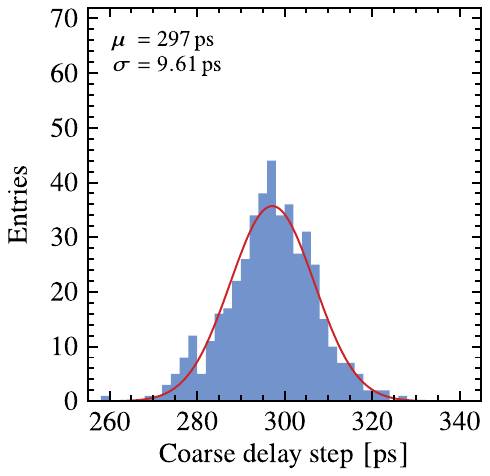}%
		\includegraphics{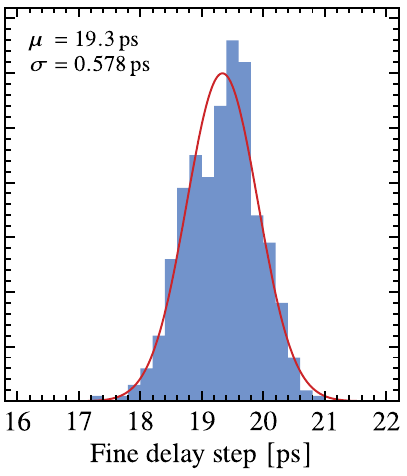}%
		\caption{Reference-clock phase shift as a function of the DLL control code for a single double-column structure (left), and the delay step distribution over all double-column structures for the coarse~(centre) and fine~(right) delay sections of the four enabled ADBs.}
		\label{fig:adbCalibration}
	\end{figure}

	\Fig{adbCalibration} shows the results for a single double-column structure, and distributions of the change in clock phase for the coarse- and fine steps over all double-column structures. A discontinuity in the clock phase shift is observed between control codes 15 and 16 which is attributed to a slight discrepancy between the coarse- and fine-element delays. The precision to which the clock phase can be controlled in the current configuration is about~\SI{20}{\pico\second}, and the total range that can be scanned with this calibration (DLL codes \numrange{0}{31}) is about \SI{600}{\pico\second}, which is sufficient for performing the analog front-end characterisation in the next section.

	\section{Analog front-end measurements} \label{sec:analogFrontEnd}
	In this section the analog front-end is characterised. First the pixel matrix is equalised in \sect{baselineEqualisation}, and then the preamplifier gain is determined in \sect{preampGain} so that the threshold can be configured in terms of signal charge. In \sect{timingPerformance} the time resolution of the analog front-end is measured as a function of signal charge and various DAC configurations are explored. Another important aspect of the timing performance is the systematic effect of signal size on the time measurement, which is the subject of \sect{timewalk}.
	
	Unless stated otherwise, the measurements in this section are performed with the DAC configuration as shown in \tab{dacConfiguration}, which is based on the values that are currently recommended by the Timepix4 manual for fast timing purposes. The bias voltage DACs are configured by tuning their outputs to the desired values using the integrated 12-bit sigma-delta ADC \cite{Casanova:2020}. The measurements are performed using three Timepix4v1 devices (labeled N2, N4, and N8), and one Timepix4v2 device (N24). Devices N2 and N8 are bonded to a \SI{300}{\micro\meter} planar silicon p-on-n sensor to provide a realistic input capacitance to the preamplifier.
	
	\begin{table}[htbp]
		\centering
		\caption{DAC configuration of the Timepix4 devices.}
		\label{tab:dacConfiguration}
		\smallskip
		\small
		\sisetup{table-align-uncertainty=true}
		\begin{tabular}{lS[table-format=3.0]}
			\multicolumn{2}{c}{Bias current DACs} \\ \toprule
			Name & {Set value}  \\ \midrule
			VBiasADC                    & 128 \\
			VBiasDAC                    &  65\makebox[0pt][l]{\textsuperscript{a}} \\
			VBiasDiscPMOS               &  89 \\
			VBiasDiscTRAFF              & 128 \\
			VBiasDiscTailNMOS           &  83 \\
			VBiasIkrum                  &   3 \\
			VBiasLevelShift             &  88 \\
			VBiasPreamp                 &  85 \\ \bottomrule
			\multicolumn{2}{@{}p{70mm}@{}}{\footnotesize{\textsuperscript{a}Increased from recommended value of 47 to improve pixel equalisation}} \\
		\end{tabular}%
		\hskip5mm%
		\begin{tabular}{lS[table-format=3.0]}
			\multicolumn{2}{c}{Bias voltage DACs} \\ \toprule
			Name & {Output [\si{\milli\volt}]}  \\ \midrule
			VCascDisc     & 550 \\
			VCascPreamp   & 750 \\
			VControlVCO\textsuperscript{b}   &   0 \\
			VFBK          & {$500/800$\textsuperscript{c}} \\
			VThreshold    & {Varied} \\
			VTpulseCoarse & {Varied} \\
			VTpulseFine   & {Varied} \\ \bottomrule
			\multicolumn{2}{@{}p{70mm}@{}}{\footnotesize{\textsuperscript{b}Only used in the Timepix4v1 devices\newline\textsuperscript{c}Electron/hole-collecting mode}} \\
			& \\
		\end{tabular}
	\end{table}

	\subsection{Baseline equalisation} \label{sec:baselineEqualisation}
	With Timepix3 the pixel-baseline equalisation is typically done by performing a threshold scan in the 10-bit counting acquisition mode for the minimum and maximum trim DAC values which control the pixel baseline. As the threshold level is increased, it moves through the noise that is superimposed on the baseline level where the number of threshold crossings peaks. The baseline level of each pixel is then taken as the peak position obtained by fitting a Gaussian to the number of threshold crossings as a function of threshold value. 
	
	Timepix4 has two frame-based counting modes that could be used to apply a similar method. However, both modes are affected by bugs which make it impractical to do so, and therefore a different method is applied. In this study the pixel-baseline equalisation of Timepix4 is performed in the data-driven 24-bit counting mode. In this mode the user sets a chip-wide counting threshold whose value is an integer multiple of 256. When the number of threshold crossings in a pixel reaches this value, it sends out a data packet and resets its counter. In this mode it is not feasible to measure a Gaussian profile of the noise since this would either lead to an impractically high data rate or a very limited counting resolution. The baseline is therefore obtained by determining two edges on either side of the noise profile of a pixel, and taking the midpoint as its baseline level. 
	
	To do so, two threshold scans are performed in opposite directions. For the first scan, all trim DACs are initialised to their lowest value, and the global threshold value is scanned in ascending order. For each threshold value the chip-wide counting threshold is set to 512, and the shutter is opened for \SI{50}{\micro\second}. When a pixel reaches a count rate of about \SI{10}{\mega\hertz} it sends out a data packet, and the equalisation routine stores the current threshold DAC value as the left noise-edge for the current trim DAC value of that pixel. Then the trim DAC is incremented by one, which increases the baseline value of that pixel, and therefore reduces its count rate until the threshold reaches the same noise edge for a second time after the scan is resumed. At the end of the scan, the left noise-edges for all pixels and trim values are known, and the scan is performed in the opposite direction to find all right noise-edges. The baseline levels are then obtained by assuming a symmetrical noise profile and taking the midpoint between the left and right noise-edges.
	
\begin{figure}[htbp]
		\centering
		\includegraphics{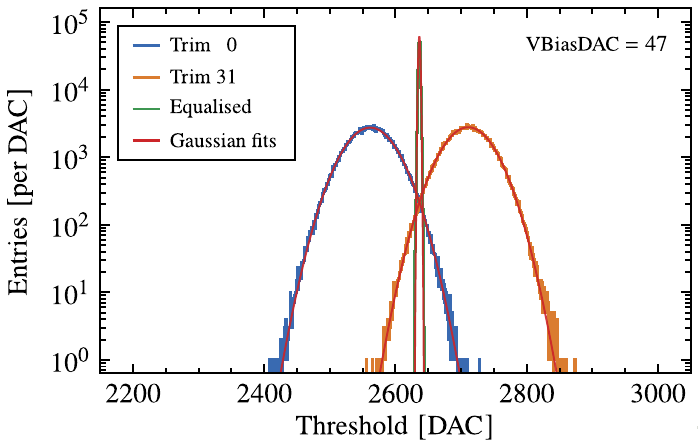}%
		\hskip5mm
		\includegraphics{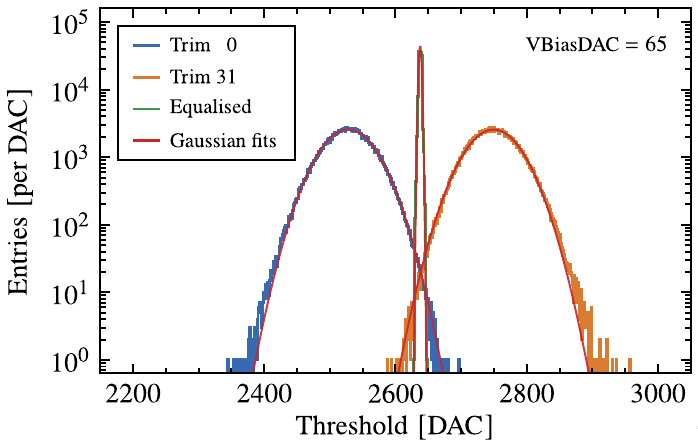}%
		\caption{Pixel-baseline distributions of N4 before and after equalisation with the bias-current DAC at its recommended value of 47 (left) and 65 (right). The device was configured in the electron-collecting mode.}
		\label{fig:equalisation}
	\end{figure}

	A target baseline value is defined as the mean baseline level of the middle trim values (15 and 16). For each pixel the trim value is chosen such that its baseline distance to the target value is minimised. \Fig{equalisation} shows the equalisation results for device N4 for two different values of the bias-current DAC that controls the pixel trim range. Increasing this DAC from its recommended value of 47 to 65 increases the trim range from about \num{150} threshold-DAC units (or \SI{83}{\milli\volt}) to 222 (\SI{123}{\milli\volt}), and reduces the fraction of pixels that cannot be equalised (because their required trim values are out-of-range) from \SI{2}{\percent} to \SI{0.1}{\percent}. Doing so, however, also increases the baseline spread from 1.43 threshold-DAC units (\SI{0.797}{\milli\volt}) to 2.01 (\SI{1.12}{\milli\volt}). The baseline spread can be expressed in electrons by taking into account the gain, which is determined in the next section. The baseline spread for the default and increased trim range are \SI{22}{\electron} and \SI{30}{\electron}, respectively. All measurements in this study are performed with the increased trim range.
	
\begin{figure}[htbp]
		\centering
		\includegraphics{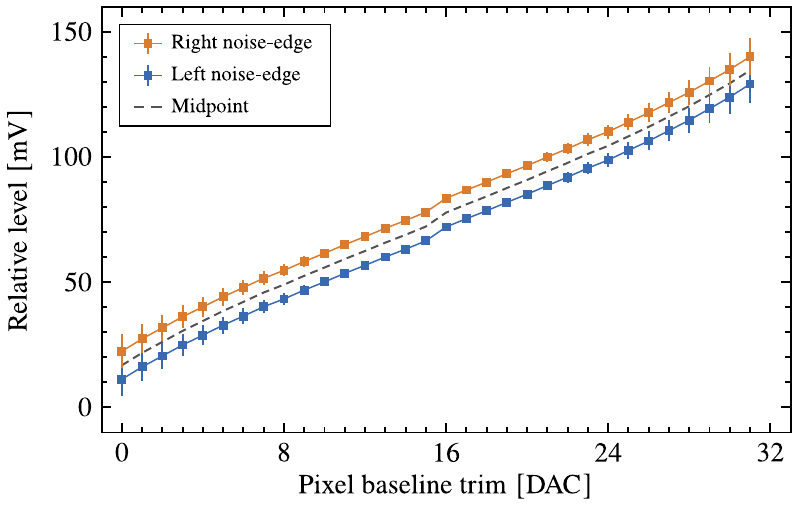}%
		\caption{Mean noise-edge levels over the pixels as a function of pixel trim value.}
		\label{fig:equalisationNoiseEdges}
	\end{figure}

	\Fig{equalisationNoiseEdges} shows the mean left and right noise-edge level as a function of the pixel trim value. Error bars indicate the one-sigma pixel-to-pixel spread. The data of each pixel are aligned at the mean midpoint value of trim values \num{15} and \num{16}. The increase in pixel-to-pixel spread towards trim values \num{0} and \num{31} reflects a divergence in the trim value dependence of the pixel baselines. The mean separation between the left and right noise-edge levels is \SI{11.3(5)}{\milli\volt}, which corresponds to about \SI{317(15)}{\electron}. This implies that the baseline spread could potentially be larger than \SI{30}{\electron}. However, in the next section the baseline spread is determined by a different method with compatible results. Nevertheless, the equalisation method may still be improved by choosing a shorter shutter time or by increasing the counting threshold in order to reduce the separation between the two noise-edges. The measurements presented in the remainder of this paper are all performed with the equalisation performed as described above.
	
	\subsection{Preamplifier gain} \label{sec:preampGain}
	A measurement of the preamplifier gain is performed in order to control the threshold in terms of signal charge. This measurement is performed by means of a threshold scan in the data-driven ToA/ToT mode with test pulses enabled in order to inject \num{1000} signals with a controlled amount of charge. To control the amount of injected charge $q$, the two test-pulse DACs are tuned such that their potential difference is equal to $q/C_{\textnormal{tp}}$, where $C_{\textnormal{tp}}$ is the nominal test-pulse coupling capacitance of \SI{3.2}{\femto\farad}. By varying the amount of injected charge, the preamplifier gain can be determined. 
	
\begin{figure}[htbp]
		\centering
		\includegraphics{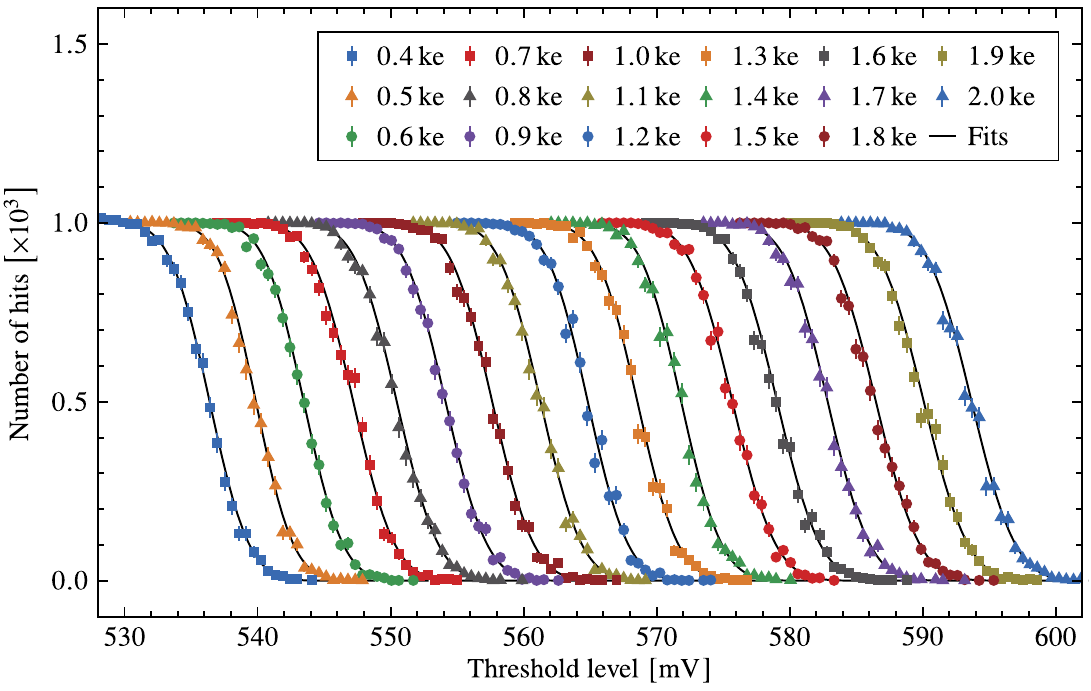}%
		\caption{Threshold scan results from a single pixel of N24 for injected test pulses ranging from \SIrange{0.6}{2}{\kilo\electron} measured in the electron-collecting mode.}
		\label{fig:gainMeasurement}
	\end{figure}

	\Fig{gainMeasurement} shows the measured number of hits as a function of threshold level in a single pixel for injected charges ranging from \SI{0.6}{\kilo\electron} to \SI{2}{\kilo\electron} in steps of \SI{100}{\electron} with the device configured in electron-collecting mode. For each scan the results are modelled as
	\begin{equation}
		\label{eq:gainThrScanFit}
		n_{\textnormal{hits}} = 
		n_0\,\mathrm{e}^{-v_{\textnormal{thr}}/\mu_v}
		+ \frac{n_{\textnormal{tp}}}{2}
		\left[
			1 - \erf\!\left(\frac{v_{\textnormal{thr}}-v_{\textnormal{peak}}}{\sqrt{2}\,\sigma_{\textnormal{v}}}\right)
		\right]\, ,
	\end{equation}
	where $v_{\textnormal{thr}}$~is the threshold level, $n_0$ and $\mu_v$~are fit parameters used to model the noise edge at the lower threshold levels, $n_{\textnormal{tp}}$~is the number of test pulses, $v_{\textnormal{peak}}$~is the peak level of the preamplifier output, and $\sigma_{\textnormal{v}}$~is the voltage noise at the preamplifier output. The relationship between $v_{\textnormal{peak}}$ and the injected charge is determined by fitting \eq{gainThrScanFit} to the measurement data. Due to the limited data rate of the slow-control readout, this measurement is performed for only 896 pixels that are located roughly along the diagonals of the bottom and top halves of the pixel matrix.
	
\begin{figure}[htbp]
		\centering
		\includegraphics{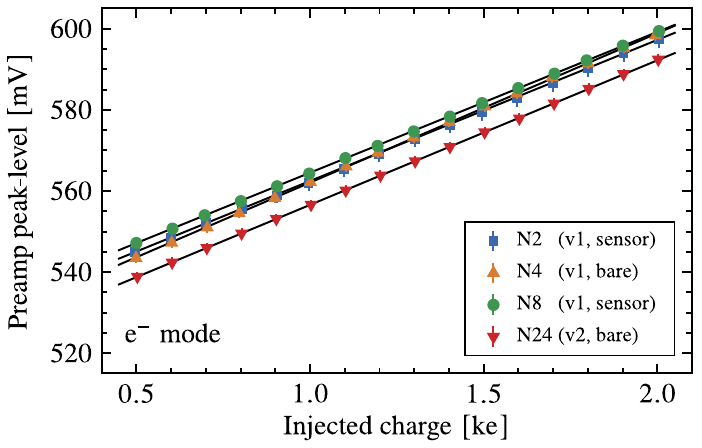}%
		\hskip5mm%
		\includegraphics{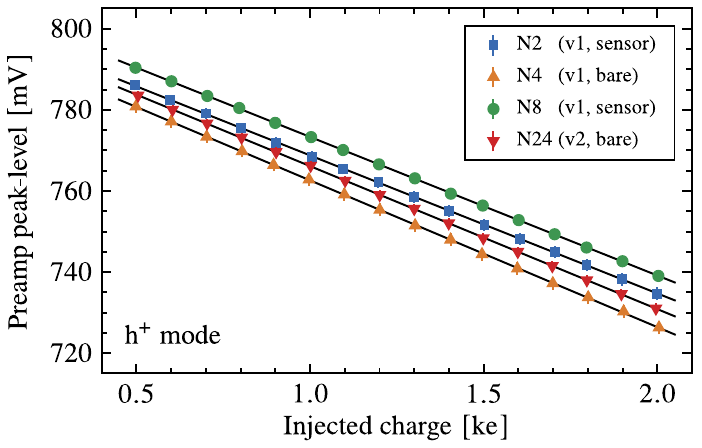}%
		\caption{Preamplifier output peak-level as a function of injected charge for all devices configured in electron-collecting mode (left) and hole-collecting mode (right). Error bars indicate the pixel-to-pixel variation.}
		\label{fig:gain}
	\end{figure}

	\Fig{gain} shows the mean preamplifier peak-level (over the pixels) as a function of injected charge for all devices and both input polarities. The gain is modelled as
	\begin{equation}
		\label{eq:gainFit}
		v_{\textnormal{peak}} = v_{\textnormal{b}} + q\, g\, ,
	\end{equation}
	where $v_{\textnormal{b}}$~the baseline level, $q$~the injected charge, and $g$~the gain. In this paper the injected charge is always quoted by its absolute value so that $q>0$, and the charge that is actually injected into the front-end is understood to be of the correct sign for the relevant polarity mode. The observed mean gain is \SI{35.0}{{\milli\volt}\per{\kilo\electron}} for collecting holes, and \SI{35.5}{{\milli\volt}\per{\kilo\electron}} for collecting electrons. The observed pixel-to-pixel RMS of the gain ranges from \SI{0.5}{\percent} to \SI{0.9}{\percent}. In this study, the linear fits shown in \fig{gain} are used to determine the threshold levels.
	
	The baseline spread is determined by fitting \eq{gainFit} for each pixel separately and calculating the standard deviation of $v_{\textnormal{b}}$. The result is divided by the mean gain to express the baseline spread in electrons. The results for all devices are presented in \tab{baselineComparison}. The baseline spread is compared to the value obtained from the equalisation as described in the previous section, and it can be seen that both measurements are in agreement. Lastly, the equivalent noise charge (ENC) is determined as $\sigma_{\textnormal{v}}/g$, and is quoted in the rightmost column. It can be seen that devices N2 and N8 have an increased ENC, which is due to the additional input capacitance provided by the bonded sensors. Both devices are fully depleted, and no significant dependence on bias potential is observed.
		
	\begin{table}[htbp]
		\centering
		\caption{Gain measurement results for all devices used in this study. The baseline spread derived from these gain measurements is compared to the value obtained from the equalisation (eq.) method. Uncertainties indicate the pixel-to-pixel variation. A reverse bias potential of \SI{100}{\volt} is applied to both sensors.}
		\label{tab:baselineComparison}
		\smallskip
		\small
		\sisetup{table-align-uncertainty=true}
		\begin{tabular}{ccS[table-format=2.0]S[table-format=2.0]S[table-format=+2.1(1)]S[table-format=2.0(1)]}
			\toprule
			       & & \multicolumn{2}{c}{Baseline spread [\si{\electron}]} & & \\ \cmidrule{3-4}
			Device & Mode & {Gain meas.} & {Eq.} & {Gain [\si{{\milli\volt}\per{\kilo\electron}}]} & {ENC [\si{\electron}]} \\ \midrule 
			\multirow{2}{*}{\makebox[0pt][l]{N2}\phantom{N24} (v1, sensor)}
				& $\textnormal{e}^-$ & 31 & 31 &  34.8(2) & 81(5)\\
				& $\textnormal{h}^+$ & 29 & 28 & -34.0(3) & 74(3)\\ \cmidrule{1-6}
			\multirow{2}{*}{\makebox[0pt][l]{N4}\phantom{N24}\makebox[0pt][l]{ (v1, bare)}\phantom{ (v1, sensor)}}
				& $\textnormal{e}^-$ & 30 & 30 &  36.8(3) & 65(2)\\
				& $\textnormal{h}^+$ & 27 & 28 & -36.2(3) & 57(2)\\ \cmidrule{1-6}
			\multirow{2}{*}{\makebox[0pt][l]{N8}\phantom{N24} (v1, sensor)}
				& $\textnormal{e}^-$ & 28 & 28 &  34.7(2) & 82(4)\\
				& $\textnormal{h}^+$ & 29 & 26 & -34.2(3) & 74(4)\\ \cmidrule{1-6}
			\multirow{2}{*}{N24\makebox[0pt][l]{ (v2, bare)}\phantom{ (v1, sensor)}}
				& $\textnormal{e}^-$ & 32 & 30 &  35.7(2) & 69(2)\\
				& $\textnormal{h}^+$ & 29 & 28 & -35.3(2) & 62(2)\\
			\bottomrule
		\end{tabular}
	\end{table}

	\subsection{Analog front-end time resolution} \label{sec:timingPerformance}
\begin{figure}[htbp]
		\centering
		\lineskip=0pt%
		\includegraphics{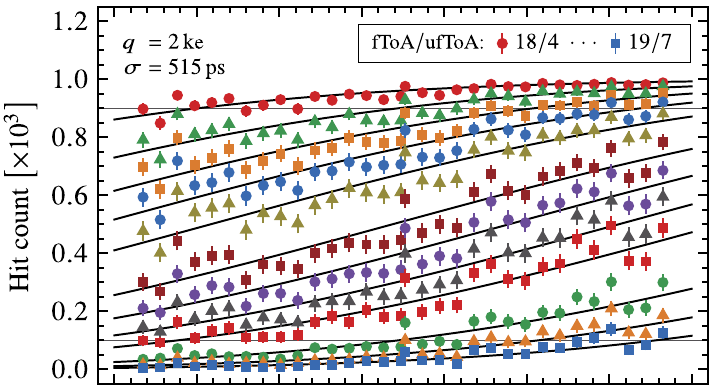}%
		\includegraphics{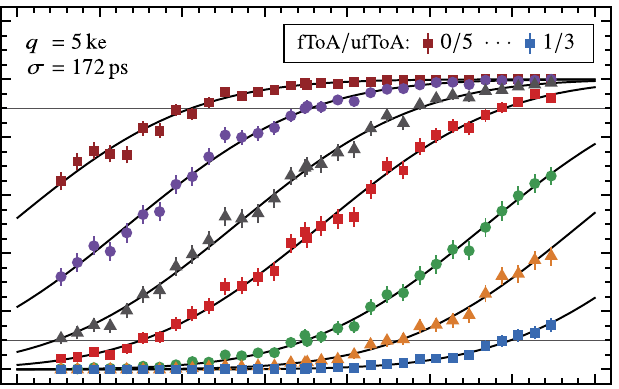}\\*[0pt]%
		\includegraphics{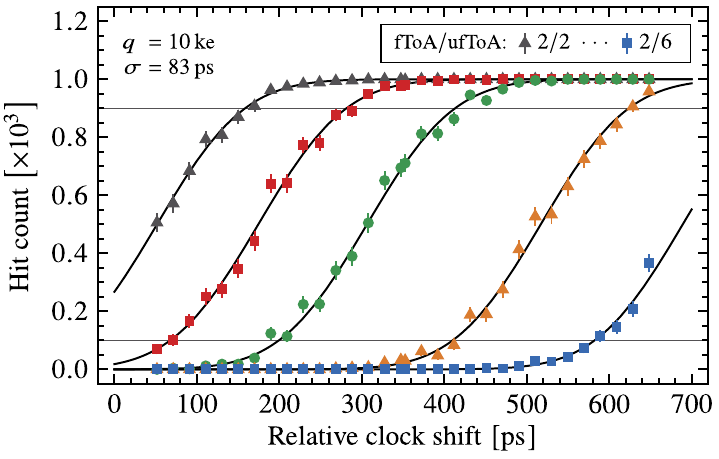}%
		\includegraphics{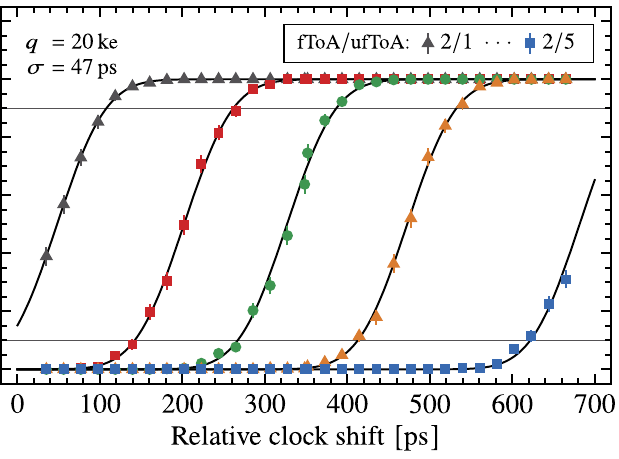}%
		\caption{Number of hits in single pixels of device N4 that arrive in or before consecutive time bins as a function of the relative clock shift for four different amounts of injected charge in the electron-collecting mode. The horizontal grey line at $y=100$ indicates the minimum number of hits that are required at the maximum clock shift for a time-bin edge to be included in the analysis, and the line at $y=900$ indicates a maximum at the minimum clock shift.}
		\label{fig:adbScan}
	\end{figure}

	The time resolution of the analog front-end (AFE) is characterised by measuring the cumulative ToA distribution of internally generated analog test pulses which inject a controlled amount of charge at the preamplifier input. The cumulative ToA distribution is measured by recording the number of hits that arrive in or before each time bin while varying the reference-clock phase using the configurable delays in the clock distribution system as described in \sect{adbCalibration}. \Fig{adbScan} shows the results of four different pixels for different amounts of charge as measured with device N4 in the electron-collecting mode. As the clock is delayed, test pulses can be seen to arrive earlier (relative to the clock) as more of them end up in time bins with larger fToA and ufToA values.
	
	For each amount of charge and pixel a simultaneous fit is performed by minimising
	\begin{equation}
		\label{eq:hitCountVsClockDelay}
		\chi^2=\sum_{i,j}
		\left[\frac{N_{ij} - f\!\left(t_j,t_i,\sigma_{\textnormal{t}}\right)}{\sigma_{ij}}\right]^2
		\, ,
	\end{equation}
	where $N_{ij}$ is the number of hits that arrived in or before the $i$-th time bin at the $j$-th step in the clock delay, $f$~is the function defined by \eq{hitCountVsTriggerDelay}, $t_j$ is the $j$-th clock delay, $t_i$ is the right-edge location of the $i$-th bin, $\sigma_{\textnormal{t}}$ is the time resolution (which is the quantity of interest), and lastly, $\sigma_{ij}$ is the statistical uncertainty that $N_{ij}$ out of $N_{\textnormal{tp}}$ test pulses arrive before the edge located at $t_i$. This uncertainty is defined by
	\begin{equation}
		\label{eq:hitCountUncertainty}
		\sigma_{ij}^2 = \frac{N_{ij}\left(N_{\textnormal{tp}}-N_{ij}\right)}{N_{\textnormal{tp}}}
		\, ,
	\end{equation}
	which is the variance of a binomial distribution: $np(1-p)$, with $n=N_{\textnormal{tp}}$ and $p=N_{ij}/N_{\textnormal{tp}}$. Note that \eq{hitCountUncertainty} requires that only data points with $0<N_{ij}<N_{\textnormal{tp}}$ are to be involved in the minimisation of \eq{hitCountVsClockDelay} to ensure that $\sigma_{ij}^2\neq0$. Furthermore, in order to ensure that the edge locations are well-defined, edges are only included in the fit when they satisfy $N_{ij}<900$ at the start of the clock-delay scan range ($j=0$) and $N_{ij}>100$ at the end of the scan range ($j=31$). Note that $\sigma_{\textnormal{t}}$ becomes large compared to the scan range for small amounts of injected charge (top left plot in \fig{adbScan}). Its value, however, is still well-constrained by the data because $N_{\textnormal{tp}}$ is fixed to \num{1000}. 
	
\begin{figure}[htbp]
		\centering
		\includegraphics{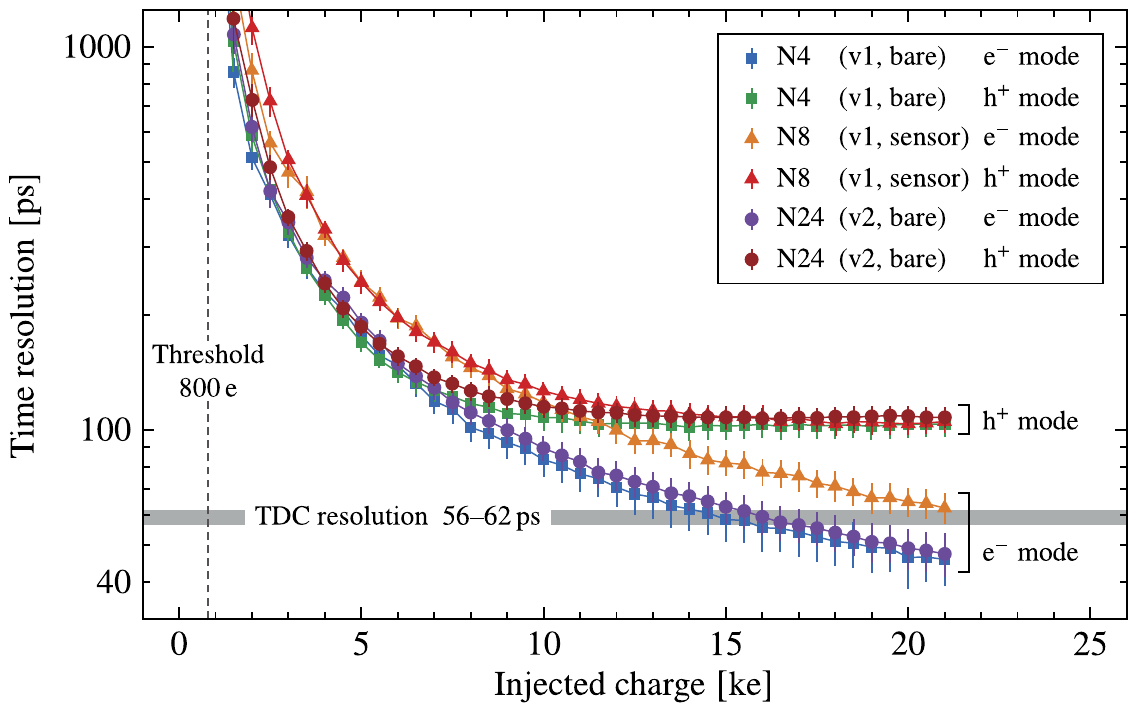}%
		\caption{Mean time resolution of the analog front-end as a function of injected charge for various devices and both polarity modes. Errors bars indicate the pixel-to-pixel variation in the time resolution. The TDC resolution is also indicated, assuming that VCO frequency variation is corrected on at least a per-VCO basis, and that time offsets between pixels are calibrated out.} 
		\label{fig:jitterVsCharge}
	\end{figure}

	\Fig{jitterVsCharge} shows the measured time resolution of the AFE as a function of injected charge. Firstly, it can be seen that the additional input capacitance from a bonded sensor has a significant impact on the time resolution. For the hole-collecting mode, the time resolution is limited due to an upper bound to the current with which the first amplifier stage can discharge its capacitive load, as was explained in \sect{timepix4PixelFrontEnd}. The AFE resolution levels off to a value of \SI{105(7)}{\pico\second}, and therefore it would dominate the total time resolution (obtained by including the TDC resolution) of \SI{122(6)} {\pico\second}. In the electron-collecting mode, however, the AFE resolution keeps improving (roughly exponentially) with signal charge, and the total time resolution is dominated by the TDC for signals larger than \SI{15}{\kilo\electron} for the bare devices. For the bonded device the break-even point is expected to be about \num{22}--\SI{23}{\kilo\electron} based on the trend.
	
\begin{figure}[htbp]
		\centering
		\includegraphics{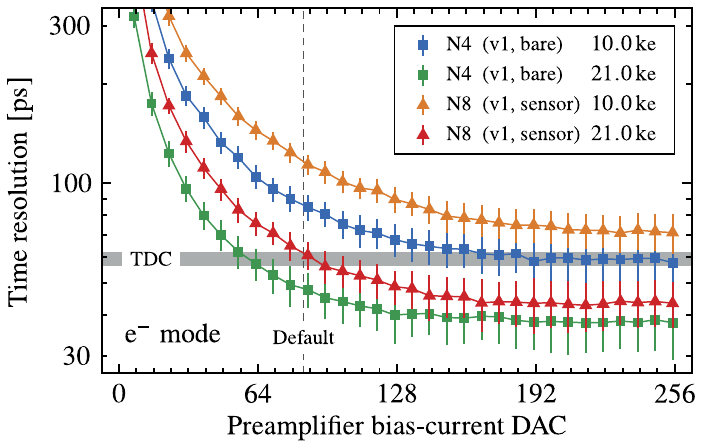}%
		\hskip5mm%
		\includegraphics{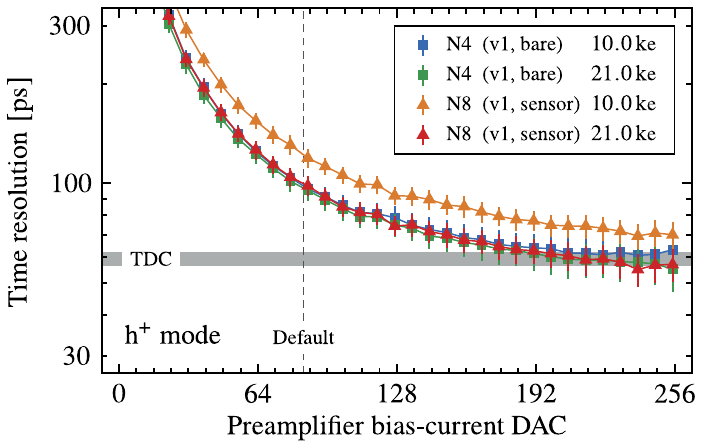}%
		\caption{Mean time resolution of the analog front-end as a function of the preamplifier bias-current DAC for various amounts of injected charge with the devices configured in electron- (left) and hole-collecting mode (right). Errors bars indicate the pixel-to-pixel variation in the time resolution.} 
		\label{fig:jitterVsPreampBias}
	\end{figure}

	The preamplifier rise-time can be decreased by increasing the preamplifier bias current, which can potentially improve the AFE time resolution. A scan of the preamplifier bias-current DAC is performed in order to study the change in time resolution, and the results are shown in \fig{jitterVsPreampBias}. In the electron-collecting mode, setting the DAC to its maximum value improves the AFE resolution by about \num{21}--\SI{38}{\percent}, depending on the signal charge and input capacitance. Increasing the bias current is more effective for lower signal charge and higher input capacitance. In the hole-collecting mode, the AFE resolution improves by about \num{35}--\SI{42}{\percent}, without a clear dependence on charge and capacitance. The improvement, however, comes at the expense of a significant increase in power consumption. In the electron-collecting mode, the total power consumption, as measured by the SPIDR4 system, increases linearly from \SI{4.1}{\watt} at the default bias current to \SI{6.0}{\watt} at the maximum value. The hole-collecting mode shows a similar increase from~\SI{3.8}{\watt} to \SI{5.6}{\watt}.
	
\begin{figure}[htbp]
		\centering
		\includegraphics{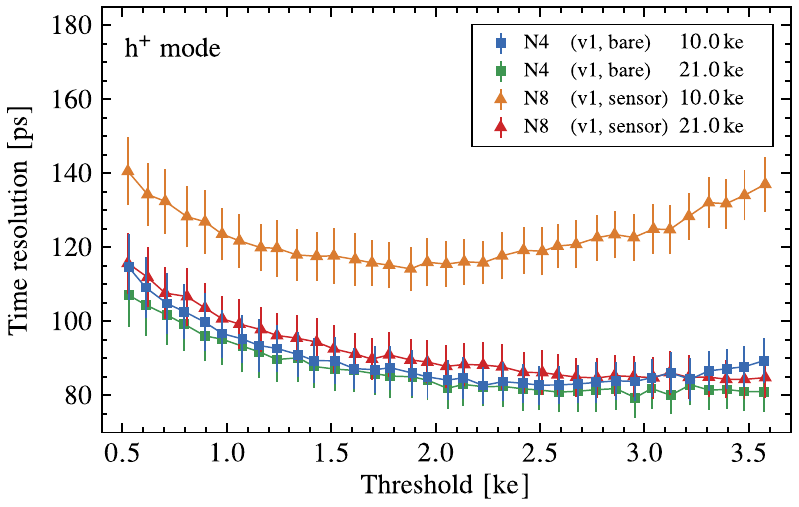}%
		\caption{Time resolution of the analog front-end as a function of threshold for various amounts of injected charge measured in the hole-collecting mode. Error bars indicate the pixel-to-pixel variation.} 
		\label{fig:jitterVsThresholdHoles}
	\end{figure}

	\Fig{jitterVsThresholdHoles} Shows the AFE time resolution as a function of threshold for the hole-collecting mode. The AFE resolution expresses a clear minimum whose location depends on both signal charge and input capacitance. For instance, the bare device (N4) has a minimum at a threshold of roughly \SI{2.5}{\kilo\electron} for a signal charge of \SI{10}{\kilo\electron} whereas the bonded device (N8), which has a larger input capacitance, has a minimum at a threshold of about \SI{2}{\kilo\electron}. Furthermore, it can be seen that the minimum of device N8 also increases with signal charge. This also seems to happen for device N4, but it is not so clear due to the limited threshold range used in the measurement. The mechanism behind this improvement can be found most likely in the slew rate of the preamplifier output. The time resolution is approximately related to the noise at the preamplifier output $\sigma_{\textnormal{v}}$ from \eq{gainThrScanFit} by
	\begin{equation}
		\sigma_{\textnormal{t}} = \frac{\sigma_{\textnormal{v}}}{dv\!/dt}
		\, ,
	\end{equation}
	where $dv\!/dt$ is the slew rate of the preamplifier output at the threshold level. By changing the threshold, the maximum slew rate can be found to optimise the time resolution.
	
\begin{figure}[htbp]
		\centering
		\includegraphics{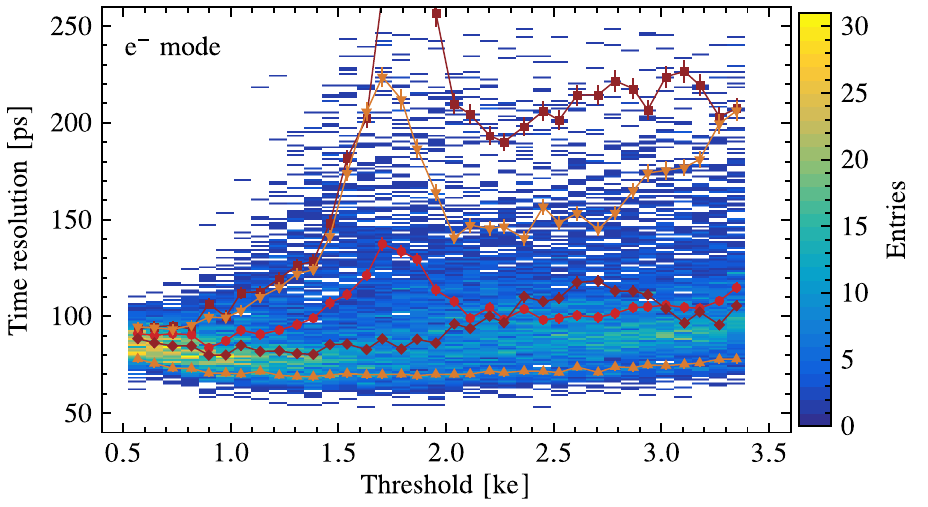}%
		\caption{Distribution of the time resolution of the analog front-end as a function of threshold measured with device N4 at a signal charge of \SI{10}{\kilo\electron} in the electron-collecting mode. The data points show the resolution of five individual pixels. For thresholds above \SI{1}{\kilo\electron} there is a significant pixel-to-pixel variation in the time resolution.} 
		\label{fig:jitterVsThresholdElectrons}
	\end{figure}

	Considering the threshold dependence in the electron-collecting mode (shown in \fig{jitterVsThresholdElectrons}), it is observed that the AFE time resolution shows a slight improvement when increasing the threshold up to a value of about \SI{1}{\kilo\electron}. At higher thresholds, however, many pixels develop a substantially worse AFE resolution, producing a tail in the distribution. On an individual pixel basis, the AFE resolution appears to peak at some pixel-specific threshold before subsequently improving again as the threshold is increased further. This effect is observed for all devices tested in this study, and the thresholds at which the AFE resolution peaks show no clear structure, do not depend on the column or row position, and also differ among the devices. The effect is currently not understood, and requires further investigation.
	
\begin{figure}[htbp]
		\centering
		\includegraphics{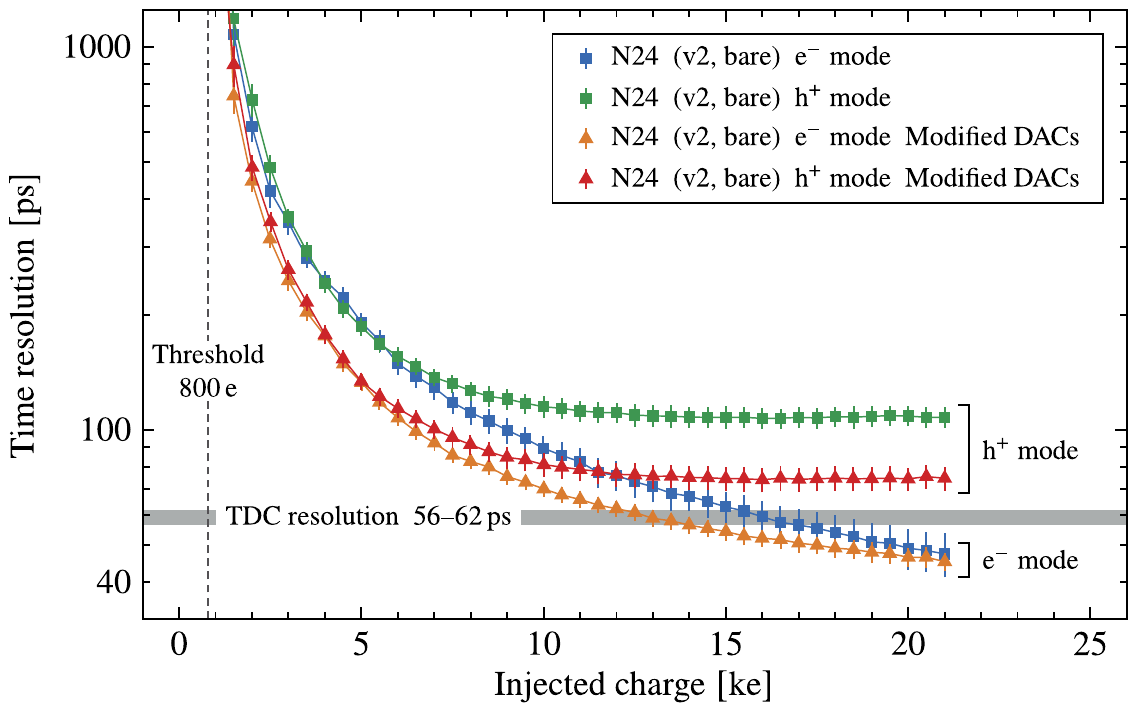}%
		\caption{Comparison of the mean analog front-end time resolution between two DAC configurations. Errors bars indicate the pixel-to-pixel variation in the time resolution. The TDC resolution is also indicated, assuming that VCO frequency variation is corrected on at least a per-VCO basis, and that time offsets between pixels are calibrated out.} 
		\label{fig:jitterVsChargeNewOp}
	\end{figure}

	Based on additional front-end simulations, a modified DAC configuration was suggested to improve the timing performance. The preamplifier bias-current DAC, which was studied above, is increased from 85 to a value of 135. In addition, the VBiasDiscPMOS DAC is increased from 89 to 135, and the VBiasDiscTRAFF is decreased from 128 to 64. This modified DAC configuration shows no significant impact on the gain and ENC measurements presented above. The improvement in AFE resolution of the Timepix4v2 device is shown in \Fig{jitterVsChargeNewOp}. In the electron-collecting mode, the modified configuration improves the AFE resolution by about \num{25} to \SI{30}{\percent} for signals up to \SI{8}{\kilo\electron}. For larger signals, the improvement decreases linearly to about \SI{3}{\percent} at \SI{21}{\kilo\electron}. The point at which the TDC starts dominating the total front-end time resolution is lowered to \SI{12}{\kilo\electron}. In the hole-collecting mode, the resolution improves by \num{25} to \SI{30}{\percent} overall, and it now levels off to a value of \SI{75(5)}{\pico\second}. The improved time resolution comes at the cost of an increase in power consumption by \SI{14}{\percent} from its baseline figure of \SI{4.1}{\watt} in the electron-collecting mode, and by \SI{20}{\percent} from its baseline figure of \SI{3.8}{\watt} in the hole-collecting mode.

	\subsection{Timewalk}\label{sec:timewalk}
	In the previous section the analog front-end time resolution was determined as a function of injected charge. In most applications the number of electron-hole pairs that are generated in the sensor material will fluctuate from one event to another---colloquially referred to as Landau fluctuations. The resulting time resolution will therefore depend on the particular charge distribution of the events. Moreover, the time resolution will be negatively affected by an additional mechanism that comes into play when combining time measurements that are performed with distinct amounts of charge: Lower-charge measurements will be systematically later than higher-charge measurements. This effect, known as \emph{timewalk}, can be corrected by means of the time-over-threshold (ToT) measurement, which is performed alongside the time-of-arrival (ToA) measurement of each hit. The ToT is a surrogate measure of the signal charge, and it can be mapped to a correction term by either a model or a lookup table. 
	
\begin{figure}[htbp]
		\centering
		\includegraphics{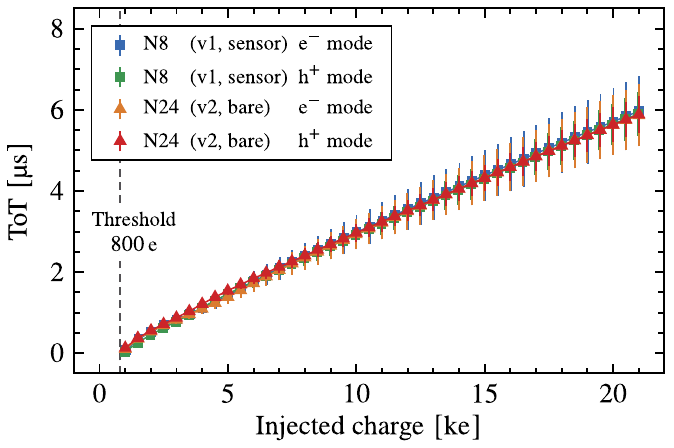}%%
		\hskip5mm%
		\includegraphics{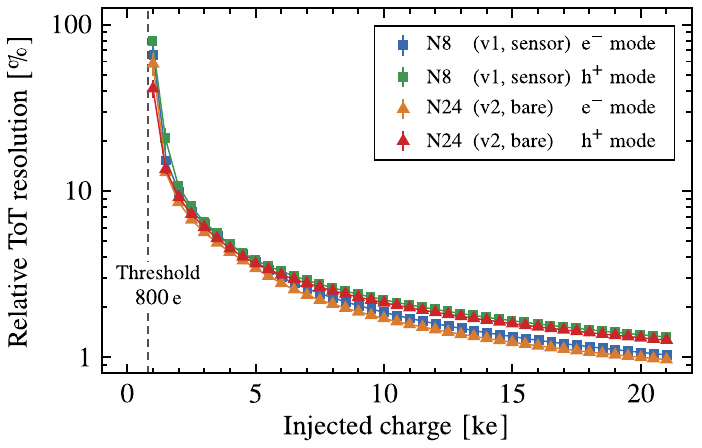}%
		\caption{Mean time over threshold (left) and the mean resolution (right) as a function of injected charge for two devices in both polarity modes. Error bars indicate the pixel-to-pixel RMS.}
		\label{fig:tot}
	\end{figure}

	The relationship between injected charge and ToT is shown in \fig{tot}. For a given amount of injected charge, a pixel-to-pixel variation in ToT is observed. This is a consequence of variation in the discharge current of the feedback capacitor onto which the signal is integrated. For injected charges larger than \SI{2}{\kilo\electron}, the pixel-to-pixel RMS of the ToT ranges from \num{11} to \SI{15}{\percent} in the electron-collecting mode, and from \num{7} to \SI{9}{\percent} in the hole-collecting mode. This effect is expected, and a per-pixel calibration is required when the ToT is used to measure signal charge or particle energy. The relative uncertainty of these measurements is determined by the relative ToT resolution, which is also shown in \fig{tot}. It improves from about \SI{10}{\percent} at a signal charge of \SI{2}{\kilo\electron} to \SI{5}{\percent} at \SI{3.5}{\kilo\electron}. For signals of \SI{21}{\kilo\electron} the relative ToT resolution is about \SI{1}{\percent}. The absolute ToT resolution ranges from about \num{40} to \SI{90}{\nano\second} depending on the polarity mode, the input capacitance, and the signal charge. The absolute ToT resolution worsens as the signal charge increases. The modified DAC settings specified in \sect{timingPerformance} have no significant impact on these results.
	
\begin{figure}[htbp]
		\centering
		\includegraphics{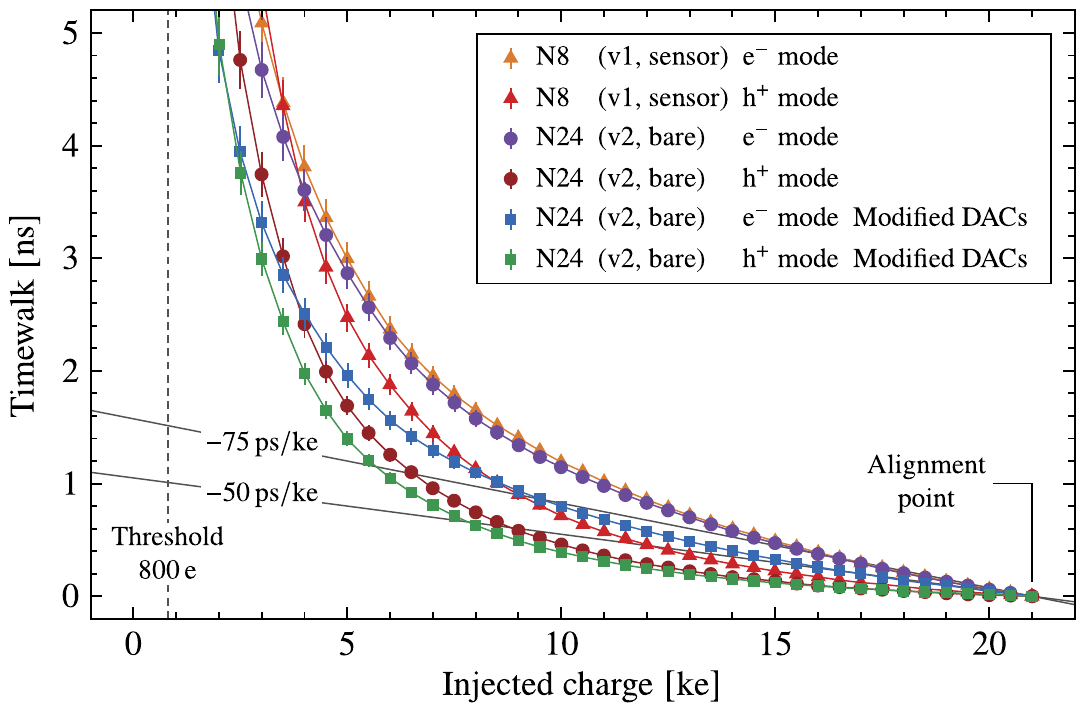}%
		\caption{Timewalk obtained by tracking the shift in time bins for various devices.} 
		\label{fig:timewalk}
	\end{figure}

	The timewalk can be extracted from the measurements of the previous section by tracking the change in time-bin locations $t_i$ from \eq{hitCountVsClockDelay} as a function of injected charge. \Fig{timewalk} shows the mean timewalk over the pixels for various devices in both collection modes. The pixels are time-aligned for a signal charge of \SI{21}{\kilo\electron}. The error bars indicate the pixel-to-pixel timewalk variation, and disappear at \SI{21}{\kilo\electron} as a result of the time alignment. The pixel-to-pixel variation can be seen to increase as the amount of injected charge decreases, which reflects a divergence in the timewalk curves of the individual pixels. In the electron-collecting mode, the timewalk behaves approximately as \SI{75}{\pico\second\per{\kilo\electron}} at a signal charge of \SI{21}{\kilo\electron}. For the modified DAC configuration, defined in \sect{timingPerformance}, the timewalk behaves as \SI{50}{\pico\second\per{\kilo\electron}}. It is also observed that the hole-collecting mode suffers from less timewalk than the electron-collecting mode, which is probably related to the difference in gain. For the hole-collecting mode, it can also be clearly seen that the bonded device (N8) has more timewalk, which may be expected as additional input capacitance typically reduces the bandwidth of an amplifier \cite{Radeka:2020}, though the effect is not so apparent in the electron-collecting mode.
	
	\section{Conclusion and outlook}\label{sec:conclusion}
	A characterisation of the Timepix4 timing performance has been performed. The pixel TDC in the digital front-end of Timepix4 has been studied using externally generated pulses that were synchronised with the \SI{40}{\mega\hertz} reference clock. Externally generated pulses have also been used to calibrate the column DLLs, which distribute the reference clock over the pixel matrix. This allowed for the characterisation of the analog front-end using internally generated analog test pulses by making it possible to control their arrival time with a step size of about \SI{20}{\pico\second}.
	
	It has been shown that the high VCO frequency has been fixed in Timepix4v2. The time bins are not completely uniform in size, but there are adjustment mechanisms available for both the PLL in the periphery and the individual VCOs in the pixel matrix that can potentially improve the time-bin uniformity. These are still to be tested. However, the non-uniformity only has a minor impact (less than \SI{1}{\percent}) on the time resolution of the TDC. It has been shown that, depending on how the VCO frequency variation is handled in the timestamp reconstruction, a TDC time resolution ranging from about \SI{58}{\pico\second} to \SI{80}{\pico\second} can be achieved. A TDC resolution of \SI{62}{\pico\second} is likely attainable with a moderate calibration effort.
	
	The pixel baselines have been equalised with a pixel-to-pixel spread of less than \SI{32}{\electron}. The fraction of pixels that cannot be equalised because their baselines deviate too much from the majority is reduced from \SI{2}{\percent} to \SI{0.1}{\percent} by increasing the bias-current DAC that controls the trim range from the default value of \num{47} to \num{65}. 
	
	Internally generated analog test pulses have been used to measure the preamplifier gain by injecting controlled amounts of charge and determining the peak level of the preamplifier response. The observed mean gain is \SI{35.5}{{\milli\volt}\per{\kilo\electron}} in the electron-collecting mode, and \SI{35.0}{{\milli\volt}\per{\kilo\electron}} in the hole-collecting mode. The ENC has been determined in the same measurement. For the bare devices an ENC ranging from \num{57} to \SI{69}{\electron{}} is observed, depending on the polarity mode. The bonded devices have a higher ENC ranging from \num{74} to \SI{82}{\electron{}} due to the additional input capacitance from the sensor.
	
	A characterisation of the analog front-end time resolution has been performed. For both polarity modes, the analog front-end time resolution shows the same signal-charge dependence up to values of about $6$--$\SI{9}{\kilo\electron}$. Beyond this point, however, the resolution levels off to \SI{105(7)}{\pico\second} in the hole-collecting mode whereas it keeps improving in the electron-collecting mode, eventually reaching values of \SI{47(7)}{\pico\second} and \SI{62(6)}{\pico\second} at an injected charge of \SI{21}{\kilo\electron} for the bare and bonded devices, respectively. The injectable charge is limited to this value by the internal DACs that define the test-pulse amplitude, and it is clear that the time-resolution still shows a tendency to improve with increasing charge. However, for these large signals the TDC will be the dominating contribution to the total front-end time resolution.

	An alternative DAC configuration has been shown to improve the analog front-end time resolution. In the electron-collecting mode, this alternative DAC configuration improves the analog front-end time resolution by \num{25} to \SI{30}{\percent} for signals up to \SI{8}{\kilo\electron}, and the point at which the TDC starts dominating the total front-end resolution is lowered from \SI{15}{\kilo\electron} to \SI{12}{\kilo\electron}. This comes at the cost of an increase in power consumption of \SI{14}{\percent}. In the hole-collecting mode, the analog front-end time resolution improves by \num{25} to \SI{30}{\percent} overall, and the limiting resolution improves from \SI{105(7)}{\pico\second} to \SI{75(5)}{\pico\second} at an increase in power consumption of \SI{20}{\percent}. For hole-collecting sensors it could also be worthwhile to experiment with the threshold level. For fixed amounts of injected charge, the analog front-end time resolution expresses a clear minimum, which can potentially improve the resolution by up to \SI{30}{\pico\second} depending on the amount of charge and the efficacy of the applied timewalk correction.
	
	Finally, to put these results into perspective, an electron-collecting version of the 3D-silicon sensor that was mentioned in the introduction is considered. For this sensor, it has been observed that a perpendicularly incident beam of minimally ionising particles generates a charge distribution that peaks at \SI{22}{\kilo\electron} and has a FWHM of about \SI{9}{\kilo\electron}~\cite{Heijhoff:2021}. Ignoring the tail towards higher charges, the signal charge can be approximated to follow a Gaussian distribution with a standard deviation of about \SI{3.8}{\kilo\electron}. Based on the measurements presented in this paper, the analog front-end time resolution for such a charge distribution is expected to have a lower limit of around \SI{62}{\pico\second}, which is the resolution of device N8 at \SI{21}{\kilo\electron}. Furthermore, a TDC resolution of \SI{62}{\pico\second} is taken into account, resulting in a lower limit for the best achievable front-end time resolution of about~\SI{88}{\pico\second}. It is also important, however, to consider the effect of timewalk. The timewalk approximately behaves as \num{50}--\SI{75}{\pico\second\per{\kilo\electron}} at a charge of \SI{21}{\kilo\electron}. Taking into account the \SI{3.8}{\kilo\electron} standard deviation of the charge distribution, it follows that timewalk alone will contribute about \num{190}--\SI{285}{\pico\second} to the total time resolution, which would make the front-end time resolution insignificant. Methods to correct for timewalk should therefore be studied in more detail when test-beam measurements are performed.

	\acknowledgments
	We express our sincere gratitude to Bas van der Heijden, Vincent van Beveren, and Henk Boterenbrood at Nikhef for the development of the SPIDR4 system and their vital support concerning its operation. This research was funded by the Dutch Research Council~(NWO).

	\bibliography{bibliography}
	
\end{document}